\documentclass[12pt,draftclsnofoot,onecolumn]{IEEEtran}

\pdfoutput=1

\usepackage{cite} 
\usepackage{xspace}
\usepackage{filecontents}
\usepackage{float}
\usepackage{pgf, tikz}
\usepackage{graphicx}
\usepackage[cmex10]{amsmath}
\usepackage{mathtools} 

\usepackage{bm}
\usepackage{amssymb} 
\usepackage{array}
\usepackage{amsfonts}
\usepackage{amsthm}   
\usepackage{enumerate} 
\usepackage{enumitem}

\usepackage{cleveref}
\crefrangeformat{equation}{(#3#1#4)\,--\,(#5#2#6)}

\newtheorem{lemma}{Lemma}

\newtheorem{remark}{Remark}
\usetikzlibrary{patterns,snakes}

\usepackage{array,algorithm,algorithmic}
\usepackage{enumerate}
\usepackage{pgfplots}
\usepackage{caption}
\usepackage{subcaption}
\usepackage{tabularx}
\usepackage{pgfplotstable}
\usepgfplotslibrary{dateplot}
\usepackage{cite}
\usepackage{url}
\usepackage{amsbsy}
\usepgfplotslibrary{patchplots}
\usepackage{xcolor}
\usepackage{tikz}
\usetikzlibrary{shadows}
\usetikzlibrary{fadings}
\usetikzlibrary{arrows}
\usetikzlibrary{shapes,snakes}
\pgfplotsset{compat=1.7}
\usetikzlibrary{patterns}
\usetikzlibrary{spy}

\usepackage{amsmath,accents}
\DeclareMathSymbol{\widehatsym}{\mathord}{largesymbols}{"62}

\DeclareMathSymbol{\widetildesym}{\mathord}{largesymbols}{"65}

\pgfplotscreateplotcyclelist{mycyclelist}{
solid, red, line width=1.0pt, every mark/.append style={solid, fill=red}, mark=square, mark size={1.2}\\%
densely dashed, blue, line width=1.0pt, every mark/.append style={solid, fill=blue}, mark=triangle, mark size={1.2}\\%
densely dotted, line width=1.0pt, green!70!black, every mark/.append style={solid, fill=green!70!black}, mark=o, mark size={1.2}\\%
densely dashdotted, line width=1.0pt, brown!70!black, every mark/.append style={solid, fill=brown!70!black}, mark=diamond, mark size={1}\\%
dotted, line width=1.0pt, orange!70!black, every mark/.append style={solid, fill=orange!70!black}, mark=x, mark size={1.3}\\%
thick, solid,  red\\%
densely dashdotted, line width=1.0pt, blue\\%
densely dotted,  line width=1.2pt, green!70!black\\%
solid,  blue, every mark/.append style={solid, fill=blue}, mark=*, mark size={1}\\%
dashed, blue, every mark/.append style={solid, fill=blue}, mark=*, mark size={1}\\%
solid,  red, every mark/.append style={solid, fill=red}, mark=square*, mark size={1}\\%
dashed, red, every mark/.append style={solid, fill=red}, mark=square*, mark size={1}\\%
solid,  green!60!black, every mark/.append style={solid, fill=green!60!black}, mark=triangle*, mark size={1.2}\\%
dashed, green!60!black, every mark/.append style={solid, fill=green!60!black}, mark=triangle*, mark size={1.2}\\%
solid,  brown!70!black, every mark/.append style={solid, fill=brown!70!black}, mark=diamond*, mark size={1.2}\\%
dashed, brown!70!black, every mark/.append style={solid, fill=brown!70!black}, mark=diamond*, mark size={1.2}\\%
}
\pgfplotscreateplotcyclelist{mycyclelist2}{
densely dashed, blue, line width=1.0pt, every mark/.append style={solid, fill=blue}, mark=triangle, mark size={1.2}\\%
densely dashdotted, line width=1.0pt, red, every mark/.append style={solid, fill=brown!70!black}, mark=x, mark size={1.2}\\%
densely dotted, line width=1.0pt, green!70!black, every mark/.append style={solid, fill=green!70!black}, mark=o, mark size={1.2}\\%
thick, solid, blue, line width=1.0pt, every mark/.append style={solid, fill=blue}, mark=triangle, mark size={1.2}\\%
thick, solid, line width=1.0pt, red, every mark/.append style={solid, fill=brown!70!black}, mark=x, mark size={1.2}\\%
thick, solid, line width=1.0pt, green!70!black, every mark/.append style={solid, fill=green!70!black}, mark=o, mark size={1.2}\\%
thick, solid,  red\\%
densely dashdotted, line width=1.0pt, blue\\%
thick, solid,  green!70!black\\%
dashed, line width=1.0pt, brown!70!black\\
densely dashdotted, blue, every mark/.append style={solid, fill=blue}, mark=square, mark size={1}\\%
solid, blue, every mark/.append style={solid, fill=blue}, mark=triangle, mark size={1.2}\\%
densely dotted, line width=0.5pt, blue, every mark/.append style={solid, fill=blue}, mark=o, mark size={1}\\%
dashed, blue, every mark/.append style={solid, fill=blue}, mark=x, mark size={1.2}\\%
thick, dashed,  green!70!black\\%
densely dashdotted, red, every mark/.append style={solid, fill=red}, mark=square, mark size={1}\\%
solid, red, every mark/.append style={solid, fill=red}, mark=triangle, mark size={1.2}\\%
densely dotted, red, every mark/.append style={solid, fill=red}, mark=o, mark size={1}\\%
dashed, red, every mark/.append style={solid, fill=red}, mark=x, mark size={1.2}\\%
}
\pgfplotscreateplotcyclelist{mycyclelist3}{
solid,  blue, every mark/.append style={solid, fill=blue}, mark=*, mark size={1}\\%
densely dashed, red, every mark/.append style={solid, fill=red}, mark=x, mark size={1.5}\\%
}

\newcommand{\black}{\color{black}}

\newcommand{\AuthorOne}{Sheeraz A. Alvi, \textit{Member,~IEEE}} 
\newcommand{\AuthorTwo}{Xiangyun Zhou, \textit{Senior~Member,~IEEE}}
\newcommand{\AuthorThree}{Salman~Durrani, \textit{Senior Member,~IEEE}}
\newcommand{\AuthorFour}{Duy T. Ngo, \textit{Member,~IEEE}}

\newcommand{\ThankOne}{Sheeraz Alvi, Xiangyun Zhou, Salman Durrani are with the Research School of Electrical, Energy and Materials Engineering, The
	Australian National University, Canberra, ACT 2601, Australia
	(emails: \{sheeraz.alvi, xiangyun.zhou, salman.durrani\}@anu.edu.au).}
\newcommand{\ThankTwo}{Duy T. Ngo is with the School of Electrical Engineering and Computing, The University of Newcastle, Callaghan, NSW 2308, Australia (e-mail: duy.ngo@newcastle.edu.au).
}

\begin{document}

\title{Sequencing and Scheduling for Multi-User Machine-Type Communication}
\author{\IEEEauthorblockN{\AuthorOne,~\AuthorTwo, \AuthorThree, and~\AuthorFour\thanks{\ThankOne}\thanks{\ThankTwo}}}
\maketitle


\begin{abstract}
%
In this paper, we propose joint sequencing and scheduling optimization for uplink machine-type communication (MTC). We consider multiple energy-constrained MTC devices that transmit data to a base station following the time division multiple access (TDMA) protocol.
%
%
Conventionally, the energy efficiency performance in TDMA is optimized through multi-user scheduling, i.e., changing the transmission block length allocated to different devices. In such a system, the sequence of devices for transmission, i.e., who transmits first and who transmits second, etc., has not been considered as it does not have any impact on the energy efficiency. In this work, we consider that data compression is performed before transmission and show that the multi-user sequencing is indeed important.
%
%
We apply three popular energy-minimization system objectives, which differ in terms of the overall system performance and fairness among the devices. We jointly optimize both multi-user sequencing and scheduling along with the compression and transmission rate control.
%
%
Our results show that multi-user sequence optimization significantly improves the energy efficiency performance of the system. Notably, it makes the  TDMA-based multi-user transmissions more likely to be feasible in the lower latency regime, and the performance gain is larger when the delay bound is stringent.
\end{abstract}

\IEEEpeerreviewmaketitle

\begin{IEEEkeywords}
Machine-type communication, sequencing, scheduling, energy consumption, data compression.
\end{IEEEkeywords}

\black

\section{Introduction}
\vspace{-0.06cm}
The Internet of Things (IoT) is largely based on the uplink communication from heterogeneous and autonomous wireless devices such as sensors, actuators which are often referred to as machine-type communication (MTC) devices \cite{fuqaha-2015,dawy-2017}. Due to their wireless and unattended operation, MTC devices are mostly battery operated and thus severely energy constrained. The energy efficient operation of these devices is therefore of pivotal importance \cite{hanzo-2017}. Specifically, wireless communication is one of the most energy-intensive operations run by the MTC devices and this calls for effective wireless solutions to prolonging the device lifetime \cite{hanzo-2017}.

The wireless MTC devices within an IoT system usually share a single uplink channel. Therein, each device first contends for channel resources and then transmits data to a central station, while satisfying stringent quality of service (QoS) requirements in terms of application specific reliability and delay. In addition, different levels of fairness among the devices are considered while minimizing the energy cost of all MTC devices (i.e., system energy cost).

In this work, the time division multiple access (TDMA) channel access mechanism is employed for the uplink MTC. The TDMA protocol allows deterministic scheduling for data transmission and other operations, such as sensing, signal detection, switching radio off, energy harvesting \cite{niyato2014}. Thus, this protocol is preferred for sensor networks \cite{cionca2008} and wirelessly powered communication networks \cite{rui2014}. It is also being promoted for automotive networks such as CAN, TTP/C, FlexRay \cite{veh-nets} due to its suitability for time-triggered and event-triggered communications. Moreover, it is also adopted for the IEEE 802.11ad standard and mmWave channel access in 5G networks \cite{mmW-5G}.

The performance of the TDMA protocol is enhanced by exploiting the multi-user diversity which occurs due to the difference in the signal power attenuation conditions of different devices. This performance enhancement is achieved by optimizing the multi-user scheduling, i.e., by changing the transmission time allocated to different devices within a frame, whilst maximizing a certain system objective such as energy efficiency, throughput. For example, a device allocated with a relatively longer transmission time can adapt the transmission rate for given channel gain, to achieve better energy efficiency and vice versa. The system objective is modelled in a particular fashion by considering the trade-off between the overall system performance and the level of fairness in terms of individual device performance. Accordingly, in a TDMA-based system, the time allocated to different MTC devices is adapted to achieve a given system objective and the devices transmit data successively in a fixed order.

From the energy efficiency perspective, there are three popular system objectives, which differ in terms of the overall system performance and fairness among the devices, i.e.,
sum, min-max, proportionally-fair energy minimization. The first objective prioritizes system performance over fairness. The second objective guarantees strict fairness among devices. The third objective strikes a balance between system performance and fairness.

For a wireless power transfer scenario, the multi-user scheduling is optimized for sum and max-min energy minimization objectives for TDMA and NOMA in \cite{rui2014} and \cite{george2016}, respectively. Similarly, the proportional-fairness objective is considered in \cite{george2017} and \cite{Guo2016} to balance between multi-user fairness and energy minimization for NOMA and TDMA, respectively. An optimal strategy is devised for TDMA in \cite{niyato2014} to balance between sum throughput and energy through multi-user scheduling. In that, the data transmission activity is controlled based on the available energy at the individual devices. For a TDMA system, in \cite{kang2014} the multi-user scheduling is optimized and the sum throughput is maximized for energy harvesting devices subject to a total time constraint. The system energy efficiency is maximized in \cite{wu2015} for TDMA systems by jointly optimizing the multi-user scheduling and transmit power subject to individual QoS requirements. To the best of our knowledge, none of the above papers investigates multi-user sequencing.

In many IoT applications, the amount of data to be transmitted is not necessarily small, resulting in a high transmission cost \cite{iomt-2015}. In this regard, data compression schemes have been proposed \cite{compressionSurvey, compression1, compression2, compression3}, which decrease the amount of data to be transmitted and thus alleviate the transmission energy cost. Typically, the energy cost of compression and transmission is around $15\%$ and $80\%$ of the total energy consumed by a sensor node, respectively, \cite{Jung-2009, raghunathan2002energy}. Unlike the transmission energy cost which linearly increases with the size of data to be transmitted, the compression energy cost has a non-linear relationship with the compression ratio \cite{tahir2013cross}. Owing to this non-linearity, blindly applying too much compression may even exceed the energy cost of transmitting raw data, thereby losing its purpose \cite{sadler2006data, sheeraz-2018, sheeraz-2018-GC}.

\vspace{-0.35cm}
\subsection{Paper Contributions}
\vspace{-0.15cm}
We consider a single-channel multi-user uplink MTC communication system, in which multiple energy-constrained MTC devices transmit data to a base station (BS) within a fixed period of time, referred to as a frame, following the TDMA protocol. The BS allocates non-overlapping frame segments, referred to as transmission blocks, to individual MTC devices. Each MTC device transmits data to the BS within its allocated transmission block. We consider that the devices apply data compression before the start of their scheduled transmission block and transmit the compressed data in the allocated transmission block. The main novelty of this work lies in the proposed multi-user sequencing, i.e., the order in which the devices are scheduled for transmission in the TDMA protocol.

Conventionally, the TDMA performance is only optimized by controlling the time allocated to different MTC devices, i.e., the length of the allocated transmission blocks. In particular, the order or sequence of devices has no significance, given the channel statistics do not change from one transmission block to the other. However, in our proposed system the sequence of allocating the devices to the transmission blocks affects the amount of time allowed for applying data compression. As such, the energy-minimization objective can be achieved by allocating MTC devices with an optimized sequence and schedule of the transmission blocks.

To this end, we propose an optimal multi-user sequencing and scheduling scheme, and a sub-optimal multi-user scheduling scheme which does not employ multi-user sequencing. A comparative performance analysis of the two proposed schemes is carried out to seek answers to the following two questions:

\begin{enumerate}

\item How much reduction in the system energy consumption can be achieved through multi-user sequence optimization for compressed transmissions?

\item In which scenarios does multi-user sequence optimization provide the most significant gains?

\end{enumerate}

\noindent Our investigation leads to the following observations and design insights:

\begin{itemize}

\item Our results show that the proposed optimal scheme outperforms the schemes without multi-user sequencing. The improvement due to multi-user sequence optimization is up to $35\%-45\%$ depending on whether the length of the transmission blocks can be optimized or not.

\item The energy efficiency gain of multi-user sequencing is most significant when the delay bound is stringent. In addition, multi-user sequence optimization makes the TDMA-based multi-user transmissions more likely to be feasible in the lower latency regime subject to the given power constraints.

\item To solve the challenging mixed-integer nonlinear program for the proposed optimal scheme, we propose transformations to arrive at an approximate convex program which can be solved with significantly lower complexity. To solve this approximate program, we develop an algorithm that iteratively converges to a Fritz John solution.

\end{itemize}

\textit{Paper Organization}: The rest of the paper is organized as follows. The system model is presented in Section~II. The proposed multi-user sequencing and scheduling problems for different system objectives are formulated in Section~III. The problem transformation and its solution strategy are presented in Section~IV. The sub-optimal scheme is presented in Section~V. Numerical results are presented in Section~VI. Finally, Section~VII concludes the paper.

\section{System model}\label{sec-system}

\begin{figure}[t]
  \centering
  \includegraphics[scale=0.950]{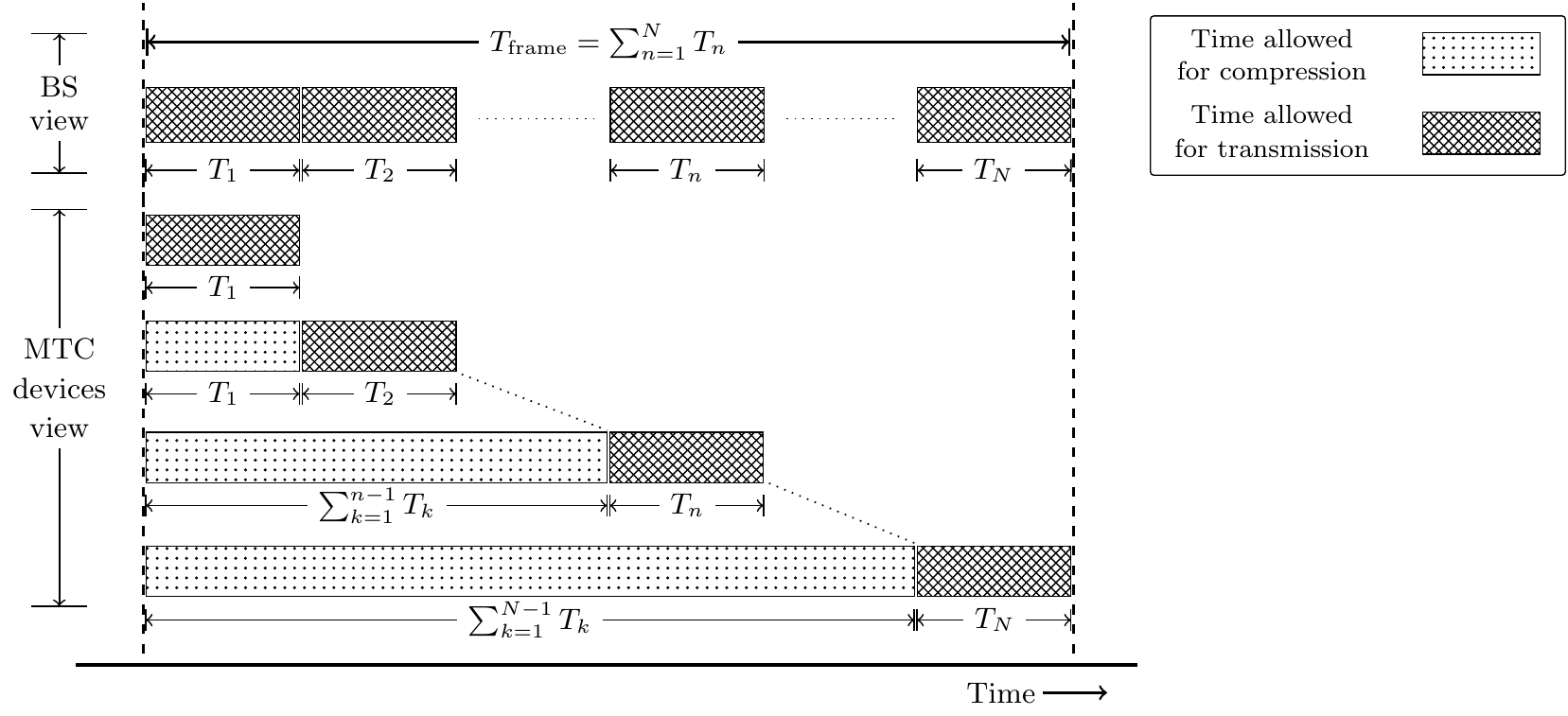}\\
  \caption{Timing diagram for the compression and transmission processes within a frame of uplink MTC. For simplicity, this figure only shows the scenario with the same block length.}
  \label{fig-timing2}
  \vspace{-0.5cm}
\end{figure}

We consider a system consisting of multiple MTC devices transmitting data packets to a BS. The devices are battery-operated and energy-constrained, whereas the BS has no energy constraint. Each device has a data packet of a specific length and the data packets of all devices need to be transmitted within a frame of length $T_\textup{frame}$ seconds. The devices employ the TDMA channel access mechanism for data transmission, as shown in Fig.~\ref{fig-timing2}. We assume perfect synchronization among devices, which is in line with recent works \cite{george2016, rui2014, george2017, Guo2016, wanchun-2016}. The devices contend for the channel by transmitting a control packet and the BS grants channel access to $N$ devices. As the details of channel contention mechanism are outside the scope of this paper, interested readers are referred to \cite{laya2014} for more information.

%
%

The BS determines the TDMA sequence and schedule and allocates non-overlapping frame segments (referred to as the transmission blocks) to individual devices. Each device is allocated a single transmission block. Both the sequence and schedule of the transmission blocks are shared with the devices by the BS before the start of the frame. Each device applies data compression before the start of its scheduled transmission block and then transmits the compressed data in the allocated transmission block, as shown in Fig.~\ref{fig-timing2}. The device allocated with the first transmission block in the frame performs both the data compression and transmission operations within its allocated transmission block. Note that the transmission block length can be different for different devices. Moreover, a device may not necessarily use all of its allocated time for compression and/or transmission.
For an energy-efficient operation, the devices are kept in power saving state when they are neither compressing nor transmitting data. We assume that the power consumed by the device in power saving state is negligible  \cite{raghunathan2002energy, Jung-2009}.

\textbf{Channel model}:
The BS and all the devices are equipped with an omnidirectional antenna. The devices are located at arbitrary distances from the BS. The distance between the $i$th device and the BS is $d_i$ meters. The channel between each device and the BS is composed of a large-scale path loss, with path loss exponent~$\alpha$, and a small-scale quasi-static frequency-flat Rayleigh fading channel. The fading channel coefficient for the $i$th device is denoted as $h_i$. All the fading channel coefficients remain unchanged over a frame and are independently and identically distributed from one frame to the next. The noise is assumed to be additive white Gaussian noise (AWGN) with zero mean and variance $\sigma^2$. The noise spectral density is given by $N_0$. The probability distribution function ($pdf$) of the instantaneous channel gain, $|h_i|^2$, is exponentially distributed~as
\begin{equation}\label{avg-avg-p}
f\big(|h_i|^2\big) \triangleq \frac{1}{\varsigma} \exp \Big( - \frac{|h_i|^2} {\varsigma} \Big), \quad |h_i|^2 \geqslant 0, \quad \forall \, i \in \{1,2,\cdot\cdot\cdot,N\},
\end{equation}
\noindent where $\varsigma$ is the scale parameter for the $pdf$. We assume that the instantaneous channel gain for each device is perfectly estimated by the BS, which is a reasonable assumption when the BS has no constraint on energy and data processing capability \cite{george2016, rui2014, george2017, liew2008}. 

\textbf{MTC device sequencing and scheduling}:
In response to the channel access requests, the BS broadcasts a control packet which contains the sequence and schedule of the device transmission blocks and the optimal compression and transmission parameters for each device. 

The frame duration, $T_\textup{frame}$, is divided into $N$ transmission blocks as
\begin{equation}
T_\textup{frame}~{=}~\sum_{n=1}^{N} T_n,
\end{equation}
\noindent where $T_n$ is the duration of $n$th transmission block.

In conventional settings, only the length of the transmission blocks affects the energy efficiency performance of a TDMA-based system. On the other hand, the sequence of the transmission block has no affect on the performance. In our case, however, the devices perform data compression before transmission. As a result, a device allocated with a later transmission block has more time to perform data compression as compared to a device allocated with an earlier transmission block. Therefore, the position of the transmission block, which depends upon the multi-user sequence, influences the achievable energy efficiency performance.

Let us define $x_{n,i}$ as:
  \begin{equation}\label{x}
 x_{n,i}  = \left\{
  \begin{array}{ll}
     1,    & \text{if the $n$th transmission block is allocated to the $i$th device}, \\
     0,    &   \textup{otherwise.}
  \end{array}
\right.
\end{equation}
As each transmission block is allocated to only one device, we have
  \begin{equation}\label{c1}
   \sum_{i=1}^{N} x_{n,i} = 1, \quad \forall \, n.
  \end{equation}
Also, each device is assigned only one transmission block, implying that
  \begin{equation}\label{c2}
    \sum_{n=1}^{N} x_{n,i} = 1, \quad \forall \, i.
  \end{equation}

\textbf{Compression}:
Before the start of its allocated transmission block, a device applies data compression on the raw data, as shown in Fig.~\ref{fig-timing2}. For the $i$th device, the $D_i$ bits of raw data is compressed into $D_{\textup{cp},i}$ bits, resulting in a compression ratio of $\frac{D_{\textup{cp},i}}{D_i}$. The \emph{compression time}, $T_{\textup{cp},i}$, is defined as the time required by the $i$th device to compress the raw data, $D_i$, into the compressed data, $D_{\textup{cp},i}$. We employ a generic non-linear compression cost model as proposed in \cite{tahir2013cross}. The parameters of this compression model can be determined off-line for a given compression algorithm using data fitting. Specifically, the performance of this compression model is validated for the JPEG and JPEG2000 compression algorithms in \cite{tahir2013cross}. Accordingly, the compression time, $T_{\textup{cp},i}$, is given as a function of compression ratio, $\frac{D_{\textup{cp},i}}{D_i}$, as
\begin{equation}\label{t-comp}
  T_{\textup{cp},i} =  \tau D_n \Big( \Big( \frac{D_i} {D_{\textup{cp},i}} \Big)^\beta - 1 \Big),
\end{equation}
\noindent where $\tau$ is the per-bit processing time and $\beta$ is a compression algorithm dependent parameter that is proportional to the compression algorithm's complexity. $\tau$ depends upon the MCU processing resources and the number of program instructions executed to process one bit of data. $\beta$ determines the time taken to achieve a given compression ratio and is calculated off-line for a specific compression algorithm and a given hardware configuration. Let $P_\textup{cp}$ be the power consumed by a device during the data compression process. $P_\textup{cp}$ is predefined and constant for a given MTC device hardware.

\textbf{Transmission}:
Once the compression process is complete, each device needs to transmit its compressed data within the allocated transmission block. The \emph{transmission time} for the $i$th device, $T_{\textup{tx},i}$, depends upon its compressed data size, $D_{\textup{cp},i}$, and its link transmission rate, $R_i$, as
\begin{equation}\label{t-comm}
  T_{\textup{tx},i} =  \frac{D_{\textup{cp},i}}{R_i}.
\end{equation}
\noindent Here, the transmission rate, $R_i$, is given as
\begin{equation}\label{r}
R_i =  B \log_2 \Big(  1 + \frac{\gamma_i}{\Gamma} \Big) ,
\end{equation}
\noindent where $B$ is the bandwidth of the considered system, $\gamma_i$ is the received signal-to-noise ratio (SNR) for the $i$th device, and $\Gamma$ characterizes the gap between the achievable rate and the channel capacity due to the use of practical modulation and coding schemes \cite{wu2016, rui2014}. The received SNR for the $i$th device, $\gamma_i$, is defined as \cite{goldsmith2005wireless}
\begin{equation}\label{snr}
 \gamma_i = \kappa \frac {P_i |h_i|^2}{\sigma^2 d_i^\alpha},
\end{equation}
\noindent where $\kappa=\big(\frac{\lambda}{4\pi}\big)^2$ is the path loss factor, $\lambda$ is the wavelength, $P_i$ is the transmit power for the $i$th device.
To compute the data transmission power cost $P_{\textup{tx},i}$ for the $i$th device, we adopt the practical model of \cite{wu2017}. The transmission power cost is composed of two components: (i)~the transmit power $P_i$ and (ii)~the static communication module circuitry power $P_\textup{o}$, which accounts for the operation of the digital-to-analog converter, frequency synthesizer, mixer, transmit filter, antenna circuits, etc. Specifically,
\begin{equation}\label{p-tx}
  P_{\textup{tx},i} =  \frac{P_i}{\mu} + P_\textup{o},
\end{equation}
\noindent where $\mu \in (0,1]$ is the drain efficiency of the power amplifier.

\section{Optimal Multi-User Sequencing and Scheduling Scheme}\label{sec-problem}

In this section, we first present the proposed multi-user sequencing and scheduling scheme and formulate the main optimization problem. Next, we present the system objectives and formulate the corresponding optimization problems. Lastly, we discuss the convexity and existence of the globally optimal solutions to the optimization problems and present our solution approach.

The MTC devices perform two main operations (i) compression and (ii) transmission, each

\noindent having individual completion time and energy cost. Recall that the device allocated with the first transmission block ($T_1$) in the frame, performs both the data compression and transmission operations within its allocated transmission block, i.e., %
\begin{equation}\label{1-device}
x_{1,i} T_{\textup{cp},i} + x_{1,i} T_{\textup{tx},i} \leqslant T_1, \quad \forall \, i.
\end{equation}
For all other devices that are allocated the remaining transmission blocks, they can apply data compression on the raw data during the period between the start of the frame and the start of its allocated transmission block, as illustrated in Fig.~\ref{fig-timing2}. This implies that the compression time for the $i$th device is upper bounded by the following constraint
\begin{equation}\label{n-device-cp}
T_{\textup{cp},i}  \leqslant  \sum_{k=1}^{n-1} T_k, \,\, \forall \, n{\geqslant}2,   \, \text{if} \, x_{n,i} =1,
\,  \text{or equivalently,}   \,
\sum_{n=2}^{N} x_{n,i} T_{\textup{cp},i}  \leqslant \sum_{n=2}^{N}  \sum_{k=1}^{n-1} x_{n,i} T_k, \,\, \forall \, i.
\end{equation}
After compression, each device transmits the compressed data within the transmission block allocated through multi-user scheduling. Accordingly, the transmission time for the $i$th device is upper bounded by the following constraint
\begin{equation}\label{n-device-tx}
T_{\textup{tx},i}  \leqslant T_n, \,\, \forall \, n{\geqslant}2,   \, \text{if} \, x_{n,i} =1,
\,  \text{or equivalently,}   \,
\sum_{n=2}^{N} x_{n,i} T_{\textup{tx},i}  \leqslant \sum_{n=2}^{N} x_{n,i} T_n, \,\, \forall \, i.
\end{equation}

Each device needs to know the following parameters for its operation: (i)~the starting time for its compression and transmission processes, (ii)~the processing time allowed for its compression and transmission processes, (iii)~the optimal compression ratio, and (iv)~the optimal transmission rate. In the considered system, the multi-user scheduling mechanism ensures there is no gap or overlap between individual transmission blocks allocated to different devices. Therefore, the starting time of both compression and transmission processes for all the devices can be determined using the starting time of the frame, the length of transmission blocks and the multi-user sequence. The starting time for compression process for all devices is equal to the starting time of the frame. The starting time for the transmission process is determined using the transmission block lengths and the multi-user sequence. Note that the transmission rate of a device is controlled through its transmit power.

\subsection{General Optimization Problem Formulation}

The main problem we address is to determine the optimal length of transmission blocks allocated to devices (i.e., scheduling), the sequence of allocated transmission blocks, the compression and transmission policies for all devices. The aim is to minimize some certain energy minimization objective (to be defined in the next subsection), under given delay and power constraints. The energy cost of the $i$th device is given as
  $  E_i = P_{\textup{cp}} T_{\textup{cp},i} +  P_{\textup{tx},i} T_{\textup{tx},i}$. Substituting the values for $T_{\textup{cp},i}$, $P_{\textup{tx},i}$ and $T_{\textup{tx},i}$ from \eqref{t-comp}, \eqref{t-comm}, \eqref{p-tx} here yields
\begin{equation}\label{En}
E_i = P_\textup{cp} \tau D_i \bigg( \Big(\frac{D_i}{D_{\textup{cp},i}} \Big)^\beta {-} 1 \bigg)   +
\frac{D_{\textup{cp},i}}{ B \log_2 \big(  1 + { \kappa \frac {P_i |h_i|^2}{\Gamma \sigma^2 d_i^\alpha}  } \big) }   \Big(  \frac{P_i}{\mu} + P_\textup{o}  \Big).
\end{equation}

The main optimization problem for the proposed scheme is formulated as follows.
\begin{subequations}\label{opt-prob}
\begin{alignat}{2}
\hspace{-1.5cm}\mathbf{P^o:} \quad &  \underset{\substack{ P_i, \, D_{\textup{cp},i}, \, T_n, \\ x_{n,i}, \, \forall \, n,i }  }    {\textup{minimize}}
& & \quad  \mathcal{E}(E_1, E_2, \cdot\cdot\cdot ,E_N)                                              \label{opt-prob-a} \\
&  \textup{subject to}
& &    \quad \sum_{n=1}^{N} T_n = T_\textup{frame},                                                 \label{opt-prob-b}\\
& & &  \quad x_{1,i} \tau D_i \bigg( \Big(\frac{D_i}{D_{\textup{cp},i}} \Big)^\beta {-} 1 \bigg)    +   x_{1,i} \frac{D_{\textup{cp},i}}{ B \log_2 \Big(  1 {+} { \kappa \frac {P_i |h_i|^2}{\Gamma \sigma^2 d_i^\alpha}  } \Big) } \leqslant T_1, \quad \forall \, i,                  \label{opt-prob-T-1}\\
& & &  \quad \sum_{n=2}^{N} x_{n,i} \tau D_i \bigg( \Big(\frac{D_i}{D_{\textup{cp},i}} \Big)^\beta {-} 1 \bigg) \leqslant \sum_{n=2}^{N} \sum_{k=1}^{n-1} x_{n,i}  T_k, \quad \forall \, i,                                                                                                  \label{opt-prob-c}\\
& & &  \quad \sum_{n=2}^{N} x_{n,i} \frac{D_{\textup{cp},i}}{ B \log_2 \Big(  1 + { \kappa \frac {P_i |h_i|^2}{\Gamma \sigma^2 d_i^\alpha}  } \Big) } \leqslant \sum_{n=2}^{N} x_{n,i} T_n,  \quad \forall \, i,      \label{opt-prob-d}\\
& & &  \quad  \sum_{i=1}^{N} x_{n,i} = 1, \qquad\qquad\quad \forall \, n,                           \label{opt-prob-e} \\
& & &  \quad  \sum_{n=1}^{N} x_{n,i} = 1, \qquad\qquad\quad \forall \, i,                           \label{opt-prob-f} \\
& & &  \quad 0 \leqslant P_i \leqslant P_{\textup{max}}, \qquad\qquad\, \forall \, i,               \label{opt-prob-g}\\
& & &  \quad D_{\textup{min},i} \leqslant D_{\textup{cp},i} \leqslant D_i, \qquad \forall \, i,     \label{opt-prob-h}\\
& & &  \quad 0 \leqslant T_n \leqslant T_\textup{frame}, \quad\qquad\:\:\, \forall \, n,            \label{opt-prob-i}\\
& & &  \quad x_{n,i} \in \{0,1\}, \qquad\qquad\;\;\; \forall \, n, i,                               \label{opt-prob-j}
\end{alignat}
\end{subequations}
\noindent where $\mathcal{E}(E_1, E_2, \cdot\cdot\cdot ,E_N)$ is the objective function imposed by the considered energy minimization strategy to be defined in the next subsection. \eqref{opt-prob-T-1}, \eqref{opt-prob-c} and \eqref{opt-prob-d} are obtained by substituting the values of compression and transmission time from \eqref{t-comp} and \eqref{t-comm}, in to inequalities \eqref{1-device}, \eqref{n-device-cp} and \eqref{n-device-tx}, respectively. $P_{\textup{max}}$ is the maximum transmit power constraint for each device. $D_{\textup{min},i}$ is the lower bound on the compressed data size for the $i$th device. Thus, the maximum compression that can be applied is given by the minimum compression ratio defined as $\frac{D_{\textup{min},i}}{D_i} \, \forall \, i$. The maximum compression ratio depends on the nature of the data and the system application. Note that a device may not fully utilize its allocated transmission block, depending upon its optimal compressed data size and/or the optimal transmission rate.

\subsection{Considered System Objectives}\label{sec-sys-OF}

In the literature, there are three popular system objectives for energy minimization, which differ in terms of the overall system performance and fairness among the MTC devices. These system objectives are (i) sum energy minimization, (ii) min-max energy minimization, and (iii) proportionally-fairness energy minimization. In the following, we discuss each of these system objectives and formulate the corresponding optimization problems. The system model under investigation allows the MTC devices to be located at various distances from the BS and experience different path attenuation. Moreover, the channel gain fluctuates independently for different devices in a given frame. This results in multi-user diversity due to the difference in the signal power attenuation conditions. The purpose behind different energy minimization objectives is to exploit this multi-user diversity while considering the trade-off between the system energy cost and the level of fairness in terms of energy cost of individual devices.

\subsubsection{Sum energy minimization}

The motivation behind the sum energy minimization objective is to prioritize the system performance over fairness among the devices. This objective attempts to achieve the maximum energy efficiency performance by fully exploiting the multi-user diversity. The strategy followed is to minimize the overall energy cost of all the devices while ensuring each device completely transmits its data within the frame duration. The energy-fairness among devices is not considered. Therein, the devices with high signal power attenuation are provided with limited system resources, thus spend more energy than other devices and vice versa.

Mathematically, the objective is to minimize the total energy cost of all the devices in the given frame, i.e., minimize $\sum_{i=1}^{N} E_i$, where $E_i$ is defined in \eqref{En}. Accordingly, for the sum energy minimization objective, \eqref{opt-prob} becomes:
\begin{equation}\label{opt-prob-sum}
\begin{aligned}
\mathbf{P^o_{SUM}:} \quad &  \underset{\substack{ P_i, \, D_{\textup{cp},i}, \, T_n, \\ x_{n,i}, \, \forall \,n,i }  }    {\textup{minimize}}
& & \sum_{i=1}^{N} E_i \big(P_i, D_{\textup{cp},i} \big) \\
& \,\,\,\, \textup{subject to}
& &    \textup{\cref{opt-prob-b,opt-prob-T-1,opt-prob-c,opt-prob-d,opt-prob-e,opt-prob-f,opt-prob-g,opt-prob-h,opt-prob-i,opt-prob-j}}.
\end{aligned}
\end{equation}

\subsubsection{Min-max energy minimization}

This system objective aims to prioritize fairness over system performance. It attempts to guarantee fairness in terms of individual device energy cost while maximizing the overall system energy efficiency performance. In particular, the maximum value of the energy spent by a device is minimized. Thereby, each device spends an equal amount of energy to transmit its data, irrespective of its signal power attenuation conditions.

The objective is to minimize the maximum energy cost among the devices in the given frame, i.e., minimize $\underset{1\leqslant i \leqslant N}{\max} \{ E_i  \}$, where $E_i$ is defined in \eqref{En}. This design strategy is Pareto-efficient \cite{Bonald2006}, i.e., the energy cost of a device cannot be further decreased without increasing the energy cost of another device. This system objective provides strict energy fairness. For the min-max energy minimization objective, \eqref{opt-prob} becomes:
\begin{equation}\label{opt-prob-equal}
\begin{aligned}
\mathbf{P^o_{MM}:} \quad &  \underset{\substack{ P_i, \, D_{\textup{cp},i}, \, T_n, \\ x_{n,i}, \, \forall \,n,i }  }    {\textup{minimize}}
& & \underset{1\leqslant i \leqslant N}{\max} \Big\{ E_i \big(P_i, D_{\textup{cp},i} \big) \Big\} \\
& \,\,\,\, \textup{subject to}
& &   \textup{\cref{opt-prob-b,opt-prob-T-1,opt-prob-c,opt-prob-d,opt-prob-e,opt-prob-f,opt-prob-g,opt-prob-h,opt-prob-i,opt-prob-j}}.
\end{aligned}
\end{equation}

\subsubsection{Proportionally-fair energy minimization}

The sum energy minimization objective prioritizes the devices with better signal power attenuation performance, thereby allocating more system resources to boost their energy efficiency. As a result, the overall system energy efficiency performance increases at the cost of energy-unfairness among the devices. On the other hand, the min-max energy minimization objective targets strict energy fairness at the cost of reduced overall system energy efficiency performance. The motivation behind proportionally-fair energy minimization objective is to strike a balance between the system energy efficiency and device energy-fairness. This objective achieves some level of fairness among devices by providing each device with a performance that is proportional to its signal power attenuation conditions. This is achieved by reducing the opportunity of the devices with low signal power attenuation, getting more share of system resources to the weak devices. More system resources are allocated to the devices when their instantaneous signal power attenuation is low relative to their own signal power attenuation statistics. Thereby, proportional-fairness is achieved without compromising much energy efficiency performance. Since the signal power attenuation fluctuates independently for different devices, this strategy effectively exploits multi-user diversity. This can be achieved by minimizing the sum of logarithmic energy cost function of the individual devices \cite{kelly1997, george2017, liew2008}, i.e., $\sum_{i=1}^{N} \log(E_i)$, where $E_i$ is defined in \eqref{En}.

For the proportionally-fair energy minimization objective,  \eqref{opt-prob} becomes:
\begin{equation}\label{opt-prob-propo}
\begin{aligned}
\mathbf{P^o_{PF}:} \quad &  \underset{\substack{ P_i, \, D_{\textup{cp},i}, \, T_n, \\ x_{n,i}, \, \forall \,n,i }  }    {\textup{minimize}}
& & \sum_{i=1}^{N} \log \Big( E_i \big(P_i, D_{\textup{cp},i} \big) \Big) \\
& \,\,\,\, \textup{subject to}
& &    \textup{\cref{opt-prob-b,opt-prob-T-1,opt-prob-c,opt-prob-d,opt-prob-e,opt-prob-f,opt-prob-g,opt-prob-h,opt-prob-i,opt-prob-j}}.
\end{aligned}
\end{equation}

\section{Problem Solution and Optimality}

The optimization problem defined in \eqref{opt-prob} is a mixed-integer nonlinear program which is non-convex in its natural form. Therefore, it is very challenging to determine the globally optimal solution or even to determine if the globally optimal solution exists \cite{bonnans2006}. In this regard, we first consider the binary variables to be deterministic (i.e., a known sequence) and transform the non-convex optimization problem into convex sub-problems using methods that preserve equivalence. This will prove that the globally optimal solution exists. Later, we propose an alternative problem modelling approach to handle the binary constraints, which targets a more computationally feasible implementation. Therein, we perform transformations on the binary variables into real variables to get an approximate convex program which can be solved with significantly lower complexity. Moreover, we propose an iterative solution approach to push these real variables towards optimal binary values, which correspond to the binary variables in \eqref{opt-prob}. These strategies are adopted for each of the considered objectives.

\subsection{Existence of Globally Optimal Solution}

For a fixed multi-user sequence \big(i.e., $\{x_{n,i}, \, \forall \, n,i\}$ is known\big), problems \eqref{opt-prob-sum}, \eqref{opt-prob-equal} and \eqref{opt-prob-propo} can be remodelled as the following sub-problems, respectively, which are still non-convex in their natural form:
\begin{equation}\label{opt-prob-sum-3}
\begin{aligned}
\hspace{-1.5cm}\mathbf{\hat{P}_{SUM}:} \quad &  \underset{P_i, \, D_{\textup{cp},i}, \, T_n, \, \forall \, n,i }    {\textup{minimize}}
& & \quad  \sum_{i=1}^{N} E_i (P_i, D_{\textup{cp},i} ) \\
& \,\,\,\, \textup{subject to}
& &  \quad  \textup{\cref{opt-prob-b,opt-prob-T-1,opt-prob-c,opt-prob-d}}, \, \textup{\cref{opt-prob-g,opt-prob-h,opt-prob-i}},
\end{aligned}
\end{equation}
\begin{equation}\label{opt-prob-equal-3}
\begin{aligned}
\hspace{-1.5cm}\mathbf{\hat{P}_{MM}:} \quad &  \underset{P_i, \, D_{\textup{cp},i}, \, T_n, \, \forall \, n,i }    {\textup{minimize}}
& & \quad  \underset{1\leqslant i \leqslant N}{\max} \Big\{ E_i (P_i, D_{\textup{cp},i} )  \Big\} \\
& \,\,\,\, \textup{subject to}
& &  \quad  \textup{\cref{opt-prob-b,opt-prob-T-1,opt-prob-c,opt-prob-d}}, \, \textup{\cref{opt-prob-g,opt-prob-h,opt-prob-i}},
\end{aligned}
\end{equation}
\begin{equation}\label{opt-prob-propo-3}
\begin{aligned}
\hspace{-1.5cm}\mathbf{\hat{P}_{PF}:} \quad &  \underset{P_i, \, D_{\textup{cp},i}, \, T_n, \, \forall \, n,i }    {\textup{minimize}}
& & \quad  \sum_{i=1}^{N} \log \Big( E_i (P_i, D_{\textup{cp},i} ) \Big) \\
& \,\,\,\, \textup{subject to}
& &  \quad  \textup{\cref{opt-prob-b,opt-prob-T-1,opt-prob-c,opt-prob-d}}, \, \textup{\cref{opt-prob-g,opt-prob-h,opt-prob-i}}.
\end{aligned}
\end{equation}

\begin{lemma}\label{lemma-1}
For a given sequence, the optimization problems \eqref{opt-prob-sum-3}, \eqref{opt-prob-equal-3} and \eqref{opt-prob-propo-3} can be transformed into corresponding equivalent convex problems. Thus, a globally optimal solution exists for each of these problems \eqref{opt-prob-sum-3}, \eqref{opt-prob-equal-3} and \eqref{opt-prob-propo-3}.
\end{lemma}
\vspace{-0.2cm}
\begin{IEEEproof}
The proof is provided in Appendix~\ref{A}.
\end{IEEEproof}

From Lemma~\ref{lemma-1}, we have the following equivalent convex problems for \eqref{opt-prob-sum-3}, \eqref{opt-prob-equal-3} and \eqref{opt-prob-propo-3}, respectively:
\vspace{-0.2cm}
\begin{subequations}\label{opt-prob-b2-sum}
\begin{alignat}{2}
\hspace{-2.2cm}\mathbf{P^{so}_{SUM}:} \quad &  \underset{  Z_i, \, D_{\textup{cp},i}, \, T_n, \,  \forall \, n,i } {\textup{minimize}}
& & \quad  \sum_{i=1}^{N} \bigg(  \tau D_i P_{\textup{cp}} \bigg( \Big(\frac{D_i}{D_{\textup{cp},i}} \Big)^\beta {-} 1 \bigg)
+ \frac{D_{\textup{cp},i} b_i} {Z_i}  \big( \exp ( Z_i ) + c_i  \big)  \bigg)    \label{b2-sum-a} \\
& \,\,\,\, \textup{subject to}
& &    \quad \eqref{opt-prob-b}, \, \eqref{opt-prob-h}, \, \eqref{opt-prob-i},   \nonumber\\
& & &  \quad x_{1,i} \tau D_i \bigg( \Big(\frac{D_i}{D_{\textup{cp},i}} \Big)^\beta {-} 1 \bigg)    +   x_{1,i} \frac{D_{\textup{cp},i}}{ B Z_i } \leqslant T_1, \quad \forall \, i,        \label{b2-T-1}\\
& & &  \quad  \sum_{n=2}^{N} x_{n,i} \tau D_i \bigg( \Big(\frac{D_i}{D_{\textup{cp},i}} \Big)^\beta {-} 1 \bigg) \leqslant \sum_{n=2}^{N} \sum_{k=1}^{n-1} x_{n,i} T_k, \quad \forall \, i, \label{b2-sum-b}\\
& & &  \quad \sum_{n=2}^{N} x_{n,i} \frac{D_{\textup{cp},i}}{ B Z_i } \leqslant \sum_{n=2}^{N} x_{n,i} T_n,  \quad \forall \, i,       \label{b2-sum-c}\\
& & &  \quad 0 \leqslant Z_i \leqslant Z_{\textup{max}}, \qquad\qquad\, \forall \, i,      \label{b1-sum-d}
\end{alignat}
\end{subequations}
\begin{equation}\label{opt-prob-b2-equal}
\begin{aligned}
\hspace{-1.75cm}\mathbf{P^{so}_{MM}:} \quad &  \underset{  Z_i, \, D_{\textup{cp},i}, \, T_n, \,  \forall \, n,i } {\textup{minimize}}
& &   \underset{1\leqslant i \leqslant N}{\max} \bigg\{ \tau D_i P_{\textup{cp}} \bigg( \Big(\frac{D_i}{D_{\textup{cp},i}} \Big)^\beta {-} 1 \bigg)
+ \frac{D_{\textup{cp},i} b_i} {Z_i}  \big( \exp ( Z_i ) + c_i  \big) \bigg\}  \\
& \,\,\,\, \textup{subject to}
& &    \eqref{opt-prob-b}, \, \eqref{opt-prob-h}, \, \eqref{opt-prob-i}, \, \textup{\cref{b2-T-1,b2-sum-b,b2-sum-c,b1-sum-d}},
\end{aligned}
\end{equation}
\begin{subequations}\label{opt-prob-b2-log}
\begin{alignat}{2}
\hspace{0.34cm}\mathbf{P^{so}_{PF}:} \quad &  \underset{  Z_i, \, V_i, \, T_n, \,  \forall \, n,i } {\textup{minimize}}
& &  \quad  \sum_{i=1}^{N} \log \bigg( \tau D_i P_{\textup{cp}}  \bigg( \frac {D^\beta_i} {\exp \big(\beta V_i\big)} {-} 1 \bigg) {+} \frac{ b_i \exp \big( V_i \big)} { Z_i } \Big( \exp\big( Z_i \big) {+} c_i  \Big) \bigg)     \label{b2-propo-a} \\
& \; \textup{subject to}
& &    \quad \eqref{opt-prob-b}, \, \eqref{opt-prob-i}, \, \eqref{b1-sum-d},  \nonumber \\
& & &  \quad x_{1,i} \tau D_i  \bigg( \frac {D^\beta_i} {\exp \big(\beta V_i\big)} - 1 \bigg) + x_{1,i} \frac{\exp \big(V_i\big) \ln(2) } { B Z_i } \leqslant T_1,  \quad \forall \, i,     \label{b2-pf-T-1}\\
& & &  \quad  \sum_{n=2}^{N} x_{n,i}\tau D_i  \bigg( \frac {D^\beta_i} {\exp \big(\beta V_i\big)} - 1 \bigg) \leqslant \sum_{n=2}^{N} \sum_{k=1}^{n-1} x_{n,i} T_k, \quad \forall \, i, \label{b2-propo-b}\\
& & &  \quad \sum_{n=2}^{N} x_{n,i}\frac{\exp \big(V_i\big) \ln(2) } { B Z_i } \leqslant \sum_{n=2}^{N} x_{n,i}T_n,   \quad \forall \, i,      \label{b2-propo-c}\\
& & &  \quad \ln(D_{\textup{min},i}) \leqslant V_i \leqslant \ln(D_i), \quad\quad\:\:\:\:\,\,\quad \forall \, i,           \label{b2-propo-d}
\end{alignat}
\end{subequations}

\noindent where $b_i{=}\frac{\Gamma \sigma^2 d^{\alpha}_i \ln(2)}{\mu B \kappa |h_i|^2}$, $c_i{=}\frac{\mu \kappa |h_i|^2 P_\textup{o}} {\Gamma \sigma^2 d^{\alpha}_i} {-} 1$, $Z_i{=}\ln \big(  1 {+} \kappa \frac {P_i |h_i|^2}{\Gamma \sigma^2 d^\alpha_i} \big)$, $Z_{\textup{max}}{=}\ln \big(  1 {+} \kappa \frac {P_{\textup{max}} |h_i|^2}{\Gamma \sigma^2 d^\alpha_i} \big)$, $V_i{=}\ln \big( D_{\textup{cp},i} \big)$.

Let us consider the exhaustive search approach for the sake of proving the existence of the globally optimal solution. Note that it is not adopted to actually solve the proposed problems. In this search approach, we first find the globally optimal solution for all possible multi-user sequence permutations for problems in \eqref{opt-prob-sum-3}. Then, we determine which multi-user sequence minimizer and its associated optimal solution gives the minimum objective value. This multi-user sequence and solution is the globally optimal solution of problem \eqref{opt-prob-sum}. The same argument applies to problems \eqref{opt-prob-equal} and \eqref{opt-prob-propo}.

\vspace{-0.25cm}
\begin{lemma}\label{lemma-2}
The minimum of the globally optimal solutions of problems \eqref{opt-prob-sum-3}, \eqref{opt-prob-equal-3} and \eqref{opt-prob-propo-3} for all possible sequences, $\{x_{n,i}, \, \forall \, n,i\}$, is the corresponding globally optimal solution of problems \eqref{opt-prob-sum}, \eqref{opt-prob-equal} and \eqref{opt-prob-propo}, respectively.
\end{lemma}
\vspace{-0.25cm}
From Lemmas 1 and 2, the globally optimal solution for each of the mixed-integer nonlinear programs in \eqref{opt-prob-sum}, \eqref{opt-prob-equal} and \eqref{opt-prob-propo}, respectively, exists and can be found.

\vspace{-0.2cm}
\subsection{Handling Binary Variables}\label{sec-binary}

In the following, we consider problem \eqref{opt-prob-sum} and first apply transformation on continuous variables as in \eqref{opt-prob-b2-sum}. Next, we transform the binary variables
into real variables to get an approximate convex program. Let us first deal with the binary variables $x_{n,i}\in \{0,1\},~\forall \,n,i$. Note that for a real variable $x_{n,i}\in [0,1]$, we have $x_{n,i} \geqslant x^{2}_{n,i},~\forall \,n,i$. To this end, we can write
\begin{equation}\label{binary-1}
x_{n,i}\in \{0,1\} \Leftrightarrow  x_{n,i} - x^{2}_{n,i}  = 0 \Leftrightarrow \big( x_{n,i}\in [0,1] ~~\&~~ x_{n,i} - x^{2}_{n,i} \leqslant 0 \big), \quad \forall \,n,i,
\end{equation}
and adopt the approach of \cite{duy31, duy30, duy29, duy2018} to rewrite the constraint function \eqref{opt-prob-j} as
\begin{equation}\label{binary-2}
\sum_{n=1}^{N} \sum_{i=1}^{N} \big( x_{n,i} - x^{2}_{n,i}  \big)  \leqslant  0,
\end{equation}
\begin{equation}\label{binary-3}
0 \leqslant x_{n,i}  \leqslant  1,  \quad \forall \,n,i.
\end{equation}
In this way, we can relax the binary variables $x_{n,i} \in \{0,1\}, \, \forall \, n,i$, in \eqref{opt-prob-sum} to real variables $x_{n,i} \in [0,1], \, \forall \, n,i$, and introduce a cost function that penalizes the objective in \eqref{opt-prob-sum} to impose $x_{n,i} = x^{2}_{n,i},~\forall \,n,i$ \cite{duy30}. Therefore, the binary to real variables transformation and the continuous variables transformation as in \eqref{opt-prob-b2-sum} leads to the following equivalent problem
\begin{equation}\label{opt-prob-real}
\begin{aligned}
\mathbf{P_\mathbb{R}:} \quad &  \underset{\substack{ Z_i, \, D_{\textup{cp},i}, \, T_n, \\ x_{n,i}, \, \forall \,n,i, }  }    {\textup{minimize}}
& & \quad  \sum_{i=1}^{N} E_i (Z_i, D_{\textup{cp},i} )  ~+~ \Lambda  \sum_{n=1}^{N} \sum_{i=1}^{N} \big( x_{n,i} - x^{2}_{n,i}  \big) \\
& \textup{subject to}
& &    \quad \eqref{opt-prob-b}, \, \eqref{opt-prob-h}, \, \eqref{opt-prob-i}, \, \textup{\cref{b2-T-1,b2-sum-b,b2-sum-c,b1-sum-d}}, \, \eqref{binary-3},
\end{aligned}
\end{equation}
\noindent where $\Lambda \geqslant 0$ is a constant penalty factor. The term $\sum_{n=1}^{N} \sum_{i=1}^{N} \big( x_{n,i} - x^{2}_{n,i}  \big)$ in \eqref{opt-prob-real} is the penalizing function on violation of the binary constraints over the energy minimization objective. Its magnitude quantifies the degree of violation from the binary constraints. $\Lambda$ embodies the cost of this violation from the binary values $x_{n,i}, \, \forall \,n,i$. The minimizer of \eqref{opt-prob-real} will satisfy the binary constraints, $x_{n,i} \in \{0,1\}, \, \forall \, n,i$, for a finite value of $\Lambda$, i.e., the penalization is exact \cite{bonnans2006}. Thus, the optimization problems defined in \eqref{opt-prob-sum} and \eqref{opt-prob-real} are equivalent, and the same optimal solution minimizes both the objective functions for a suitable value of the penalty factor \cite{duy30}.

The non-negative term $\sum_{n=1}^{N} \sum_{i=1}^{N} \big( x_{n,i} - x^{2}_{n,i}  \big)$ in \eqref{opt-prob-real} decreases to 0 as $\Lambda \rightarrow + \infty$. Ideally, we need this term to be zero, and for that we would have to derive the optimal value of the penalty factor, $\Lambda^*$. For practical computational feasibility, let us introduce a numerical tolerance level such that it is acceptable to have $\sum_{n=1}^{N} \sum_{i=1}^{N} \big( x_{n,i} - x^{2}_{n,i}  \big) < \epsilon $, where $\epsilon$ is very small and $\Lambda$ is sufficiently large. Following \cite{duy31} and \cite{duy30}, in our numerical experiments we found $\Lambda \geqslant 200$ is large enough to satisfy a tolerance level of $\epsilon = 10^{-6}$ such that $\sum_{n=1}^{N} \sum_{i=1}^{N} \big( x_{n,i} -  x^{2}_{n,i}  \big) \leqslant \epsilon$.

Note that the penalty function in \eqref{opt-prob-real} is non-convex in $x_{n,i},~\forall \,n,i,$. Consider a non-convex quadratic function $g(x) \triangleq x-x^2,$ where $x \in [0,1]$. If we apply the first-order Taylor series expansion at a given point $x^{(j)} \in [0,1]$, we can obtain the convex lower bound on $g(x)$ as \cite{duy2018}\vspace{-0.15cm}
\begin{equation}\label{bi-convex-1}
x \big( 1 - 2 x^{(j)} \big) + \big( x^{(j)} \big)^2 \leqslant x - x^2.\vspace{-0.2cm}
\end{equation}
Similarly, the convex lower bound on the penalty function can also be given as\vspace{-0.1cm}
\begin{equation}\label{bi-convex-2}
 \sum_{n=1}^{N} \sum_{i=1}^{N} \Big( x_{n,i} \big( 1 - 2 x^{(j)}_{n,i} \big) + \big( x^{(j)}_{n,i} \big)^2  \Big)
  \leqslant
 \sum_{n=1}^{N} \sum_{i=1}^{N} \big( x_{n,i} - x^{2}_{n,i}  \big).\vspace{-0.08cm}
\end{equation}

Accordingly, for a given point $x^{(j)}_{n,i} \in [0,1]$, the global upper bound minimization for problem \eqref{opt-prob-real} is given as\vspace{-0.15cm}
\begin{equation}\label{opt-prob-real-convex}
\begin{aligned}
\mathbf{P^{UB}_\mathbb{R}:} \quad &  \underset{\substack{ Z_i, \, D_{\textup{cp},i}, \, T_n, \\ x_{n,i}, \, \forall \,n,i, }  }    {\textup{minimize}}
& & \quad  \sum_{i=1}^{N} E_i (Z_i, D_{\textup{cp},i} )  ~+~ \Lambda  \sum_{n=1}^{N} \sum_{i=1}^{N} \Big( x_{n,i} \big( 1 - 2 x^{(j)}_{n,i} \big) + \big( x^{(j)}_{n,i} \big)^2  \Big)  \\
& \textup{subject to}
& &    \quad \eqref{opt-prob-b}, \, \eqref{opt-prob-h}, \, \eqref{opt-prob-i}, \, \textup{\cref{b2-T-1,b2-sum-b,b2-sum-c,b1-sum-d}}, \, \eqref{binary-3}.
\end{aligned}
\end{equation}

\vspace{-0.25cm}
\subsection{Solution Approach}

Algorithm~1 outlines the steps to find the solution to the nonconvex problem \eqref{opt-prob-sum} by iteratively solving a number of convex problem \eqref{opt-prob-real-convex}. In the first iteration, $j=1$, problem \eqref{opt-prob-real-convex} is solved using the initially guessed points, $x^{(j)}_{n,i}, \, \forall \, n,i$. The solution for the $j$th iteration $x^*_{n,i} , \, \forall \, n,i$ is used as an initial point for next iteration $j+1$. This process is repeated until convergence is achieved. The final solution yields the optimal parameters for multi-user sequencing and scheduling and compression and transmission rates for problem \eqref{opt-prob-sum}  due to its equivalence to problem \eqref{opt-prob-real-convex}.

\setlength{\textfloatsep}{14pt}
\begin{algorithm}[t]
\caption{Iterative Approach for Multi-User Sequencing and Scheduling Optimization}
\begin{algorithmic}[1]
\STATE \textbf{Initialization}: Set iteration count $j=0$. Set initial point for $x^{(j)}_{n,i}=0.5, \, \forall \, n,i$.  Select a reasonably high penalty value $\Lambda=200$ and low tolerance value $\epsilon= 10^{-6}$.
\WHILE{$\sum_{n=1}^{N} \sum_{i=1}^{N} \big( x^{(j)}_{n,i} - \big( {x^{(j)}_{n,i}}\big)^2 \big) \geqslant \epsilon$}
   \STATE Solve  \eqref{opt-prob-real-convex} using point $x^{(j)}_{n,i}, \, \forall \, n,i$ and get solution parameters $Z^*_i, \, D^*_{\textup{cp},i}, \, T^*_n, \, x^*_{n,i}, \, \forall \, n,i$.
   \STATE Update point $x^{(j+1)}_{n,i} = x^*_{n,i},  \, \forall \, n,i$
   \STATE Update iteration count $j = j + 1$
\ENDWHILE
\end{algorithmic}
\end{algorithm}

\vspace{-0.15cm}
\begin{lemma}\label{lemma-3}
Algorithm 1 converges to a Fritz John solution of problem \eqref{opt-prob-sum}.
\end{lemma}
\vspace{-0.25cm}
\begin{IEEEproof}
The proof is based on \cite{duy2018}, \cite{duy-36}, \cite{duy-25} and it is provided in Appendix~\ref{B}.
\end{IEEEproof}
\vspace{0.2cm}
Problem \eqref{opt-prob-real-convex}  is solved in each iteration of Algorithm 1 with a polynomial computational complexity in the number of variables and constraints. Accordingly, \eqref{opt-prob-real-convex}  can be transformed into an equivalent optimization problem such that it contains $\text{n}_\textup{l} {=} (5N{+}1)$ linear constraints, $\text{n}_\textup{p} {=} (3N)$ posynomial constraints, and $\text{n}_\textup{b} {=} (N^2)$ real-valued scalar decision variables. Thus, solving \eqref{opt-prob-real-convex} requires a complexity of
$\mathcal{O} \big(\sqrt{ \text{n}_\textup{l} {+} \text{n}_\textup{p} }   [\text{n}_\textup{l} {+} \text{n}_\textup{p} {+} \text{n}_\textup{b} ]  \text{n}^2_\textup{b} \big)$ \cite{duy29}.

It is noted that Algorithm~1 can be straightforwardly modified to solve problems  \eqref{opt-prob-equal} and \eqref{opt-prob-propo} by first applying the continuous variables transformation as in \eqref{opt-prob-b2-equal} and \eqref{opt-prob-b2-log}, respectively, and then following the same steps provided in Section~\ref{sec-binary} for the binary variables transformation.


\vspace{-0.2cm}
\section{Sub-optimal Multi-User Scheduling Scheme}\label{sec-bench}

In this section, we present a sub-optimal multi-user scheduling scheme which employs compression but, unlike the optimal scheme, does not consider multi-user sequencing. In Section~\ref{sec-results}, we will analyze the relative energy efficiency performance of the proposed optimal scheme and the sub-optimal scheme in a multi-user uplink communication system. For this reason, we will keep our focus on jointly optimizing multi-user sequencing and scheduling and compression ratio. Three optimization problems are formulated for the sub-optimal scheme considering the three different system objectives previously defined in Section~\ref{sec-sys-OF}.

For the sub-optimal scheme, the multi-user sequence is fixed and unchanged from one frame to the next. However, the transmission block length of any device is flexible and can be optimized. In this scheme, the transmission rate, compression ratio, and the transmission block length (multi-user scheduling) are jointly optimized for the given energy minimization objective for a fixed multi-user sequence \big(i.e.,~$\{x_{n,i}, \, \forall \, n,i\}$ is known\big). For the considered system objectives, the optimization problems for the sub-optimal scheduling scheme are given in \eqref{opt-prob-b2-sum}, \eqref{opt-prob-b2-equal} and \eqref{opt-prob-b2-log}.

\vspace{-0.1cm}
\section{Numerical Results}\label{sec-results}

This section presents the numerical results to illustrate the performance of the proposed scheme. Unless specified otherwise, the values for the parameters shown in Table \ref{para-table} are adopted.
\vspace{-0.85cm}
\begin{remark}\label{remark-solu}
Algorithm~1 is implemented in AMPL \textup{\cite{ampl}}, which is popular for modelling scheduling problems\footnote{To cross-check we compared AMPL-Couenne with CVX-sedumi/SDPT3 in geometric programming (GP) mode. The CVX-GP mode requires additional transformations on \eqref{opt-prob-real-convex} for compliance. Nevertheless, the results match up to 6 decimal points.}. A model for the proposed problem is developed in the AMPL environment and the Couenne (convex over and under envelopes for nonlinear estimation) solver \textup{\cite{couenne, couenne2}} is used to solve the problem. The Couenne solver guarantees the globally optimal solution if such a solution exists and we have already proved its existence by Lemmas 1 and 2.
\end{remark}
\vspace{-0.15cm}
Let us define system energy cost as the total energy cost of all the devices, i.e., $\sum_{i=1}^{N} E_{i}$. Moreover, the energy efficiency gain, $\mathbb{G}_\textup{ee}$, provided by a given scheme $A$ over scheme $B$ be defined as the percentage decrease in the system energy cost of scheme $B$, $\sum_{i=1}^{N} E_{i,B}$, in comparison to the system energy cost of scheme $A$, $\sum_{i=1}^{N} E_{i,A}$, and it is given as
\begin{equation}
\mathbb{G}_\textup{ee} = \frac{\sum_{i=1}^{N} E_{i,B} -  \sum_{i=1}^{N} E_{i,A}}{\sum_{i=1}^{N} E_{i,B}}.
\end{equation}

It should be noted that the relative performance of the sum, min-max, and proportionally-fair energy minimization objectives has been well studied in previous studies and thus it is not the focus of this paper. Our focus is rather to evaluate the performance of the proposed joint optimization of multi-user sequencing and scheduling scheme.

\begin{table}[]
\centering
\caption{System Parameter Values.}
\label{para-table}
\begin{tabular}{|l|c|c||l|c|c|} \hline
\textbf{Name}                                & \textbf{Sym.}  & \textbf{Value}    & \textbf{Name}   & \textbf{Sym.}  & \textbf{Value}                   \\ \hline
Amplifier's drain efficiency          & $\mu$            & 0.35              & Max. transmit power    & $P_\textup{max}$ & 0 dB                             \\
Scale parameter for channel gain      & $\varsigma$      & 1                 & Wavelength             & $\lambda$        & 0.333 m                          \\
Compression processing power          & $P_\textup{cp}$  & 24 mW             & No. of devices         & $N$              & 5                                \\
Comm. module circuitry power          & $P_\textup{o}$   & 82.5 mW           & Bandwidth              & $B$              & 1 MHz                            \\
Practical modulation power gap        & $\Gamma$         & 9.8 dB            & Packet size            & $D_i$            & \{310,500,100,80,200\} kbits     \\
Minimum compression ratio             & $\frac{D_{\textup{min},i}}{D_i}$ & 0.4   & Distance           & $d_i$            & \{40,15,31,49,22\} m             \\
Per-bit processing time               & $\tau$           & 7.5 ns/b          & Noise spectral density & $N_0$            & $-$174 dBm                         \\
Compression cost parameter            & $\beta$          & 5                 & Pathloss exponent      & $\alpha$         & 4                                \\ \hline
\end{tabular}
\vspace{-0.25cm}
\end{table}

To the best of our knowledge, the recent works \cite{george2017, Guo2016, wu2015, rui2014} are the most relevant to our proposed scheme. Although the system models in these works are based on wireless power transfer, the underlying multi-user scheduling and transmission rate policy designs are similar to our considered system. In this regard, we adopt the multi-user scheduling and transmission rate design policies proposed by these schemes for our considered system model except that data compression and multi-user sequencing are not employed. Moreover, when our considered system is applied, the design problems proposed in \cite{george2017, Guo2016, wu2015, rui2014} can equivalently be represented by the following benchmark scheme.

\textit{Benchmark scheme}: The multi-user sequence is fixed but the transmission block length of any device is flexible and can be optimized. This scheme does not include data compression and multi-user sequencing optimization. The transmission rate and the transmission block length (scheduling) are jointly optimized for the given energy minimization objective for a fixed sequence and without employing data compression. For comparison with the proposed scheme, the same energy minimization objectives are considered. The corresponding optimization problems for the benchmark scheme are given in Appendix~\ref{C}. The strategy followed to optimize the multi-user scheduling and device transmission rate policies for this benchmark scheme is essentially the same as in the state of the art \cite{george2017, Guo2016, wu2015, rui2014}.

\setlength{\textfloatsep}{30pt}
\begin{figure}
\hspace{3.5cm}
\begin{subfigure}{.5\textwidth}
    \begin{tikzpicture}[spy using outlines={ chamfered rectangle, magnification=2.0, width=1.25cm, height=2.0cm,, connect spies}]
    \begin{axis}[
     height=7.0cm, width=8.5cm,
     legend cell align=left,
     legend style={inner xsep=1pt, inner ysep=1pt,at={(0.95,0.80)},anchor=east,font=\tiny, legend columns=1, draw, fill},
     cycle list name = mycyclelist,
     mark repeat={1},
     grid=both,
     label style={font=\small},
     xtick pos=left,
     ytick pos=left,
     xmin=50, xmax=150,  xtick={50,70,90,110,130,150},
     xlabel= Frame duration: $T_\textup{frame}$ (ms),
     ylabel = System energy cost: $\sum_{n=1}^{N} E_n$ (mW),
     ylabel style={yshift=-0.5ex},
     ymin=0, ymax=250, ytick={0,50,100,150,200,250,300},
     ticklabel style={
        /pgf/number format/fixed,
        /pgf/number format/precision=5
     }
     ]
        \addplot table[x expr={\thisrow{x}*1000}, y expr={\thisrow{y}*1000}] {proposed_sum.txt};
        \addplot table[x expr={\thisrow{x}*1000}, y expr={\thisrow{y}*1000}] {bench_sum_CP_noSeq.txt};
        \addplot table[x expr={\thisrow{x}*1000}, y expr={\thisrow{y}*1000}] {bench_sum_noCP_noSeq.txt};
    \legend{Optimal scheme $\big(\mathbf{P^o_{SUM}}\big)$, Sub-optimal scheme $\big(\mathbf{P^{so}_{SUM}}\big)$, Benchmark scheme $\big(\mathbf{P^{b}_{SUM}}\big)$}
    \end{axis}
        \end{tikzpicture}
\caption{Sum-energy minimization}
\end{subfigure}
\vspace{0.25cm}
\\
\begin{subfigure}{.5\textwidth}
\hspace{-0.35cm}
\begin{tikzpicture}[spy using outlines={ chamfered rectangle, magnification=2.0, width=1.25cm, height=2.0cm,, connect spies}]
    \begin{axis}[
     height=7.0cm, width=8.5cm,
     legend cell align=left,
     legend style={inner xsep=1pt, inner ysep=1pt,at={(0.95,0.80)},anchor=east,font=\tiny, legend columns=1, draw, fill},
     cycle list name = mycyclelist,
     mark repeat={1},
     grid=both,
     label style={font=\small},
     xtick pos=left,
     ytick pos=left,
     xmin=50, xmax=150,  xtick={50,70,90,110,130,150},
     xlabel= Frame duration: $T_\textup{frame}$ (ms),
     ylabel = System energy cost: $\sum_{n=1}^{N} E_n$ (mW),
     ylabel style={yshift=-0.5ex},
     ymin=0, ymax=250, ytick={0,50,100,150,200,250,300},
     ticklabel style={
        /pgf/number format/fixed,
        /pgf/number format/precision=5
     }
     ]
        \addplot table[x expr={\thisrow{x}*1000}, y expr={\thisrow{y}*1000}] {proposed_equal.txt};
        \addplot table[x expr={\thisrow{x}*1000}, y expr={\thisrow{y}*1000}] {bench_equal_CP_noSeq.txt};
        \addplot table[x expr={\thisrow{x}*1000}, y expr={\thisrow{y}*1000}] {bench_equal_noCP_noSeq.txt};
    \legend{Optimal scheme $\big(\mathbf{P^o_{MM}}\big)$, Sub-optimal scheme $\big(\mathbf{P^{so}_{MM}}\big)$, Benchmark scheme $\big(\mathbf{P^{b}_{MM}}\big)$}
    \end{axis}
        \end{tikzpicture}
\caption{Min-max energy minimization}
\end{subfigure}
\,
\begin{subfigure}{.5\textwidth}
    \begin{tikzpicture}[spy using outlines={ chamfered rectangle, magnification=2.0, width=1.25cm, height=2.0cm,, connect spies}]
    \begin{axis}[
     height=7.0cm, width=8.5cm,
     legend cell align=left,
     legend style={inner xsep=1pt, inner ysep=1pt,at={(0.95,0.80)},anchor=east,font=\tiny, legend columns=1, draw, fill},
     cycle list name = mycyclelist,
     mark repeat={1},
     grid=both,
     label style={font=\small},
     xtick pos=left,
     ytick pos=left,
     xmin=50, xmax=150,  xtick={50,70,90,110,130,150},
     xlabel= Frame duration: $T_\textup{frame}$ (ms),
     ylabel = {},
     ylabel style={yshift=-0.5ex},
     ymin=0, ymax=250, ytick={0,50,100,150,200,250,300},
     ticklabel style={
        /pgf/number format/fixed,
        /pgf/number format/precision=5
     }
     ]
        \addplot table[x expr={\thisrow{x}*1000}, y expr={\thisrow{y}*1000}] {proposed_log.txt};
        \addplot table[x expr={\thisrow{x}*1000}, y expr={\thisrow{y}*1000}] {bench_log_CP_noSeq.txt};
        \addplot table[x expr={\thisrow{x}*1000}, y expr={\thisrow{y}*1000}] {bench_log_noCP_noSeq.txt};
    \legend{Optimal scheme $\big(\mathbf{P^o_{PF}}\big)$, Sub-optimal scheme $\big(\mathbf{P^{so}_{PF}}\big)$, Benchmark scheme $\big(\mathbf{P^{b}_{PF}}\big)$}
    \end{axis}
        \end{tikzpicture}
\caption{Proportionally-fair energy minimization}
\end{subfigure}
\vspace{-0.15cm}
\caption{System energy cost under given power constraints and system objectives.}
\label{optimal-vs-sub-1-and-bench}
\vspace{-0.5cm}
\end{figure}
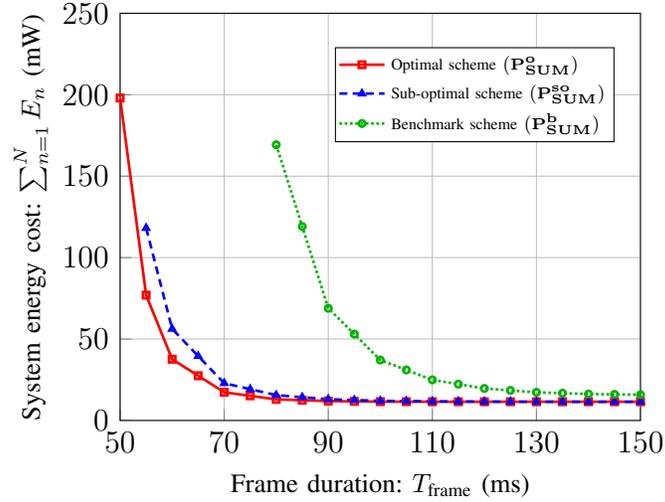
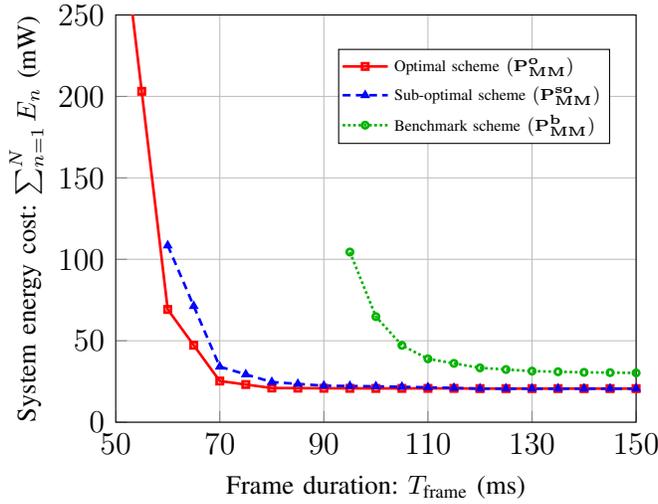
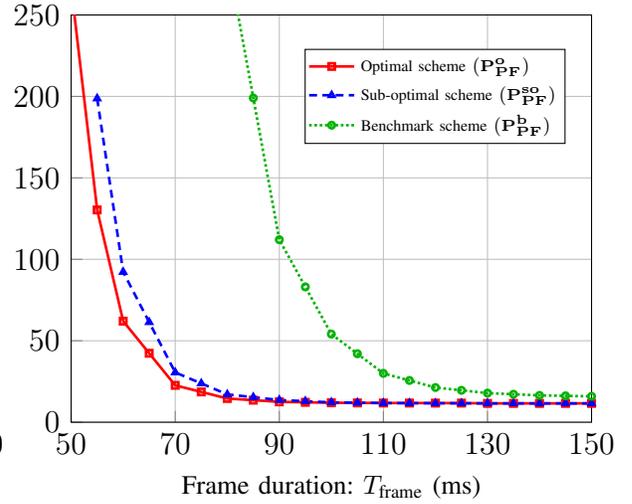

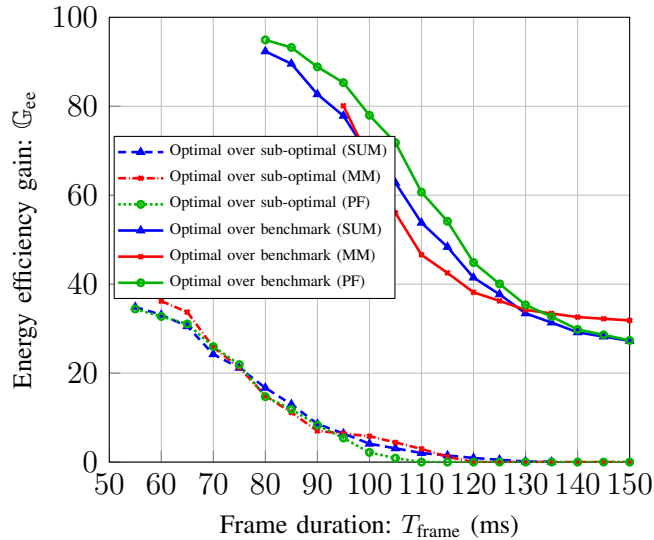
\begin{figure}
\centering
    \begin{tikzpicture}[spy using outlines={ chamfered rectangle, magnification=2.0, width=1.25cm, height=2.0cm,, connect spies}]
    \begin{axis}[
     height=7.5cm, width=8.5cm,
     legend cell align=left,
     legend style={inner xsep=1pt, inner ysep=1pt,at={(0.55,0.55)},anchor=east,font=\tiny, legend columns=1, draw, fill}, 
     cycle list name = mycyclelist2,
     mark repeat={1},
     grid=both,
     label style={font=\small},
     xtick pos=left,
     ytick pos=left,
     xmin=50, xmax=150,  xtick={50,60,70,80,90,100,110,120,130,140,150},
     xlabel= Frame duration: $T_\textup{frame}$ (ms),
     ylabel= Energy efficiency gain: $\mathbb{G}_\textup{ee}$ ,
     ylabel style={yshift=-0.5ex},
     ymin=0, ymax=100, 
     ticklabel style={
        /pgf/number format/fixed,
        /pgf/number format/precision=5
     }
     ]

        \addplot table[x expr={\thisrow{x}*1000}, y expr={\thisrow{y}*100}] {proposed_over_bench_2_sum.txt};
        \addplot table[x expr={\thisrow{x}*1000}, y expr={\thisrow{y}*100}] {proposed_over_bench_2_equal.txt};
        \addplot table[x expr={\thisrow{x}*1000}, y expr={\thisrow{y}*100}] {proposed_over_bench_2_log.txt};
        \addplot table[x expr={\thisrow{x}*1000}, y expr={\thisrow{y}*100}] {proposed_over_bench_1_sum.txt};
        \addplot table[x expr={\thisrow{x}*1000}, y expr={\thisrow{y}*100}] {proposed_over_bench_1_equal.txt};
        \addplot table[x expr={\thisrow{x}*1000}, y expr={\thisrow{y}*100}] {proposed_over_bench_1_log.txt};
    \legend{Optimal over sub-optimal (SUM), Optimal over sub-optimal (MM), Optimal over sub-optimal (PF), Optimal over benchmark (SUM), Optimal over benchmark (MM), Optimal over benchmark (PF) }
    \end{axis}
        \end{tikzpicture}
\caption{Energy efficiency gain performance vs. frame duration for the proposed optimal scheme over the sub-optimal scheme and benchmark scheme, for the considered system objectives.}
\label{optimal-vs-sub-1-and-bench-gain}
\vspace{-0.5cm}
\end{figure}

\vspace{-0.1cm}
\subsection{Validation}

In this subsection, we carry out a comparative analysis of the proposed scheme with the benchmark scheme (which represents existing state-of-the-art work). Fig. \ref{optimal-vs-sub-1-and-bench} plots the system energy cost, $\sum_{i=1}^{N} E_i$, versus the frame duration, $T_\textup{frame}$, for the system parameters in Table~\ref{para-table}. The system energy cost is plotted with the proposed optimal scheme and benchmark scheme for the sum, min-max, and proportionally-fair energy minimization objectives in Fig.~\ref{optimal-vs-sub-1-and-bench}. The energy efficiency gain, $\mathbb{G}_\textup{ee}$, provided by the proposed optimal scheme over benchmark scheme is plotted in Fig.~\ref{optimal-vs-sub-1-and-bench-gain}. The performance for the sub-optimal scheme is also shown in Figs.~\ref{optimal-vs-sub-1-and-bench} and \ref{optimal-vs-sub-1-and-bench-gain} which we will discuss later. It can be seen from Fig.~\ref{optimal-vs-sub-1-and-bench-gain} that the gains are almost the same irrespective of the considered energy minimization objective. In the following, we will discuss system performance for the sum energy minimization objective. Similar conclusions can be drawn for the min-max and proportionally-fair energy minimization objectives.

When compared with the benchmark scheme, the proposed optimal scheme exhibits significant performance superiority. This shows that employing both multi-user sequence and compression optimization provides notable gains in the energy efficiency, specifically in the lower latency regime. For the sum energy minimization objective, Fig.~\ref{optimal-vs-sub-1-and-bench-gain} shows that the gain is comparatively significant (between $27\%$ to $92\%$), for the considered range of delay when the system is feasible for benchmark scheme (between $150$ ms to $80$ ms). For the benchmark scheme, the device energy cost is reduced by adapting the minimum required transmit power level under given channel conditions. However, reducing transmission rate through transmit power only helps up to a certain level and any further reduction does not improve energy efficiency. Hence, in general, it is not optimal to transmit at the lowest transmission rate. Note that for the proposed optimal scheme the lower bound delay has a much smaller value as compared to the benchmark scheme.

\vspace{-0.11cm}
\subsection{Impact of Multi-User Sequencing}

To illustrate the advantage of the proposed joint multi-user sequencing, we consider the proposed optimal scheme and sub-optimal scheme for comparative analysis. In both schemes, the multi-user scheduling and compression are optimized. However, they differ in an important aspect that the multi-sequencing is employed by the proposed optimal scheme and not by sub-optimal scheme, which uses a fixed multi-user sequence.

From Fig.~\ref{optimal-vs-sub-1-and-bench}, the proposed optimal multi-user sequencing and scheduling scheme clearly outperforms the sub-optimal scheme. Intuitively, it was expected that the multi-user sequencing will always provide non-negative gains. However, the gains are notable, between $13\%$ to $35\%$, for the considered range of delay when the system is feasible for the sub-optimal scheme as shown in Fig.~\ref{optimal-vs-sub-1-and-bench-gain}, for the sum energy minimization objective. Also, in the lower latency regime the gains are significantly high, between $85$ ms to $55$ ms. Thus, for a less stringent delay constraint, employing the multi-user sequencing will not pay off. At the same time, it can be conclude that the data compression provides significant gains for all sorts of delay constraints.

In addition, when the proposed optimal scheme is employed, the  TDMA-based multi-user transmissions is more likely to be feasible in the lower latency regime subject to the given power constraints. That is, the proposed optimal scheme can support much stringent delay requirements, under given maximum transmit power constraints, as compared to the sub-optimal scheme for the same system parameters. Note that in Fig.~\ref{optimal-vs-sub-1-and-bench}, the system energy cost flattens out as the delay is increased further from a specific value for both schemes. The reason is that for both proposed schemes, with joint data compression and transmission rate strategy, there exists a lower bound on the device energy cost.

\subsection{Impact of Scheduling Flexibility}

To illustrate the impact of multi-user scheduling on the overall system performance and multi-user sequencing, we consider a simple scenario in which each transmission block allocated to individual device is fixed and equal in length, i.e., the frame duration, $T_\textup{frame}$, is divided into $N$ equal transmission blocks and assigned to $N$ devices. For this scenario, we consider two cases:

\begin{itemize}
  \item \textit{Case 1}: the multi-user sequence is fixed,
  \item \textit{Case 2}: the multi-user sequence can be optimized.
\end{itemize}

\noindent In both cases, the transmission rate and compression ratio are optimized for each device under a given energy minimization objective. The corresponding optimization problems for both these cases are given in Appendices~\ref{D} and \ref{E}, respectively.

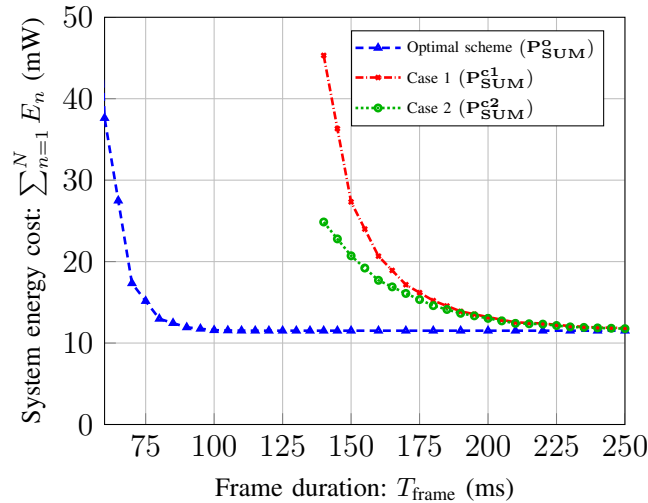
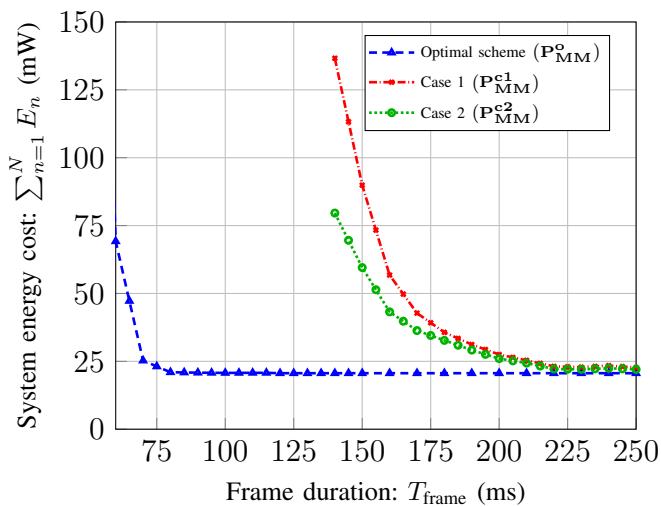
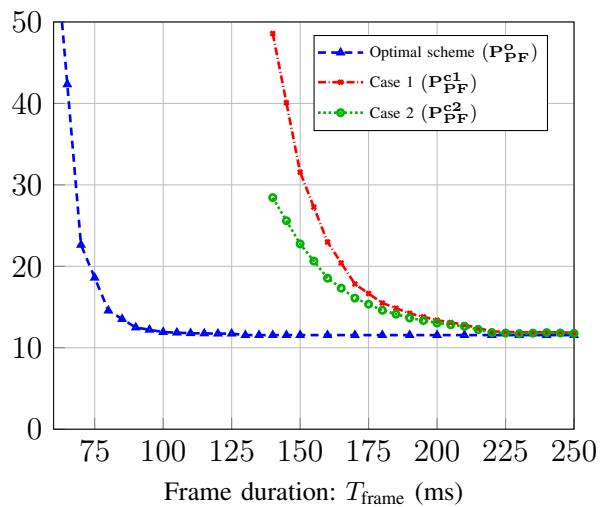
\begin{figure}
\hspace{3.5cm}
\begin{subfigure}{.5\textwidth}
    \begin{tikzpicture}[spy using outlines={ chamfered rectangle, magnification=2.0, width=1.25cm, height=2.0cm,, connect spies}]
    \begin{axis}[
     height=7.0cm, width=8.5cm,
     legend cell align=left,
     legend style={inner xsep=1pt, inner ysep=1pt,at={(0.96,0.85)},anchor=east,font=\tiny, legend columns=1, draw, fill},
     cycle list name = mycyclelist2,
     mark repeat={1},
     grid=both,
     label style={font=\small},
     xtick pos=left,
     ytick pos=left,
     xmin=60, xmax=250, xtick={50,75,100,125,150,175,200,225,250},
     xlabel = Frame duration: $T_\textup{frame}$ (ms),
     ylabel = System energy cost: $\sum_{n=1}^{N} E_n$ (mW),
     ylabel style={yshift=-0.5ex},
     ymin=0, ymax=50, ytick={0,10,20,30,40,50,60},
     ticklabel style={
        /pgf/number format/fixed,
        /pgf/number format/precision=5
     }
     ]
        \addplot table[x expr={\thisrow{x}*1000}, y expr={\thisrow{y}*1000}] {proposed_sum.txt};
        \addplot table[x expr={\thisrow{x}*1000}, y expr={\thisrow{y}*1000}] {bench_sum_CP_Seq_noSch.txt};
        \addplot table[x expr={\thisrow{x}*1000}, y expr={\thisrow{y}*1000}] {bench_sum_CP_noSeq_noSch.txt};
    \legend{Optimal scheme $\big(\mathbf{P^{o}_{SUM}}\big)$, Case 1 $\big(\mathbf{P^{c1}_{SUM}}\big)$, Case 2 $\big(\mathbf{P^{c2}_{SUM}}\big)$}
    \end{axis}
        \end{tikzpicture}
\caption{Sum-energy minimization}
\end{subfigure}
\vspace{0.25cm}
\\
\begin{subfigure}{.5\textwidth}
\hspace{-0.35cm}
    \begin{tikzpicture}[spy using outlines={ chamfered rectangle, magnification=2.0, width=1.25cm, height=2.0cm,, connect spies}]
    \begin{axis}[
     height=7.0cm, width=8.5cm,
     legend cell align=left,
     legend style={inner xsep=1pt, inner ysep=1pt,at={(0.95,0.85)},anchor=east,font=\tiny, legend columns=1, draw, fill},
     cycle list name = mycyclelist2,
     mark repeat={1},
     grid=both,
     label style={font=\small},
     xtick pos=left,
     ytick pos=left,
     xmin=60, xmax=250, xtick={50,75,100,125,150,175,200,225,250},
     xlabel = Frame duration: $T_\textup{frame}$ (ms),
     ylabel = System energy cost: $\sum_{n=1}^{N} E_n$ (mW),
     ylabel style={yshift=-0.5ex},
     ymin=0, ymax=150, ytick={0,25,50,75,100,125,150},
     ticklabel style={
        /pgf/number format/fixed,
        /pgf/number format/precision=5
     }
     ]
        \addplot table[x expr={\thisrow{x}*1000}, y expr={\thisrow{y}*1000}] {proposed_equal.txt};
        \addplot table[x expr={\thisrow{x}*1000}, y expr={\thisrow{y}*1000}] {bench_equal_CP_Seq_noSch.txt};
        \addplot table[x expr={\thisrow{x}*1000}, y expr={\thisrow{y}*1000}] {bench_equal_CP_noSeq_noSch.txt};
   \legend{Optimal scheme $\big(\mathbf{P^{o}_{MM}}\big)$, Case 1 $\big(\mathbf{P^{c1}_{MM}}\big)$, Case 2 $\big(\mathbf{P^{c2}_{MM}}\big)$}
    \end{axis}
        \end{tikzpicture}
\caption{Min-max energy minimization}
\end{subfigure}
\,
\begin{subfigure}{.5\textwidth}
    \begin{tikzpicture}[spy using outlines={ chamfered rectangle, magnification=2.0, width=1.25cm, height=2.0cm,, connect spies}]
    \begin{axis}[
     height=7.0cm, width=8.5cm,
     legend cell align=left,
     legend style={inner xsep=1pt, inner ysep=1pt,at={(0.95,0.85)},anchor=east,font=\tiny, legend columns=1, draw, fill},
     cycle list name = mycyclelist2,
     mark repeat={1},
     grid=both,
     label style={font=\small},
     xtick pos=left,
     ytick pos=left,
     xmin=60, xmax=250, xtick={50,75,100,125,150,175,200,225,250},
     xlabel = Frame duration: $T_\textup{frame}$ (ms),
     ylabel = {}, 
     ylabel style={yshift=-0.5ex},
     ymin=0, ymax=50, ytick={0,10,20,30,40,50,60},
     ticklabel style={
        /pgf/number format/fixed,
        /pgf/number format/precision=5
     }
     ]
        \addplot table[x expr={\thisrow{x}*1000}, y expr={\thisrow{y}*1000}] {proposed_log.txt};
        \addplot table[x expr={\thisrow{x}*1000}, y expr={\thisrow{y}*1000}] {bench_log_CP_Seq_noSch.txt};
        \addplot table[x expr={\thisrow{x}*1000}, y expr={\thisrow{y}*1000}] {bench_log_CP_noSeq_noSch.txt};
    \legend{Optimal scheme $\big(\mathbf{P^{o}_{PF}}\big)$, Case 1 $\big(\mathbf{P^{c1}_{PF}}\big)$, Case 2 $\big(\mathbf{P^{c2}_{PF}}\big)$}
    \end{axis}
        \end{tikzpicture}
\caption{Proportionally-fair energy minimization}
\end{subfigure}
\vspace{-0.15cm}
\caption{System energy cost under given power constraints and system objectives.}
\vspace{-0.5cm}
\label{sub-2-vs-sub-3}
\end{figure}

Fig. \ref{sub-2-vs-sub-3} plots the system energy cost, $\sum_{i=1}^{N} E_i$, versus the frame duration, $T_\textup{frame}$, for the system parameters in Table~\ref{para-table}. The system energy cost is plotted with the proposed optimal scheme and Case~1 and Case~2 for the sum, min-max, and proportionally-fair energy minimization objectives in Fig.~\ref{sub-2-vs-sub-3}. As can be seen, both Case~1 and Case~2 perform similar. However, when the delay is stringent Case~2 performs significantly better than Case~1 due to multi-user sequencing. The energy efficiency gain is between $11\%$ to $45\%$, for the considered range of delay from $165$ ms to $140$ ms for the sum energy minimization objective.

When compared with the optimal scheme, a large performance degradation is observed for the considered scenario when the transmission block lengths are fixed. Thus, optimizing the multi-user sequencing provides a large performance gain even in a restricted scheduling scenario. In addition, multi-user scheduling flexibility when combined with multi-user sequencing has a significant impact on the overall system performance. Similar conclusions can be drawn for the min-max and proportionally-fair energy minimization objectives.

\section{Conclusion}
\vspace{-0.1cm}
In this paper, we have investigated the joint optimization of sequencing and scheduling in a multi-user uplink machine-type communication scenario, considering adaptive compression and transmission rate control design. The energy efficiency performance is evaluated for three energy-minimization system objectives, which differ in terms of the overall system performance and fairness among the devices.
Our results have showed that the proposed optimal scheme outperforms the schemes without multi-user sequencing. The improvement in energy efficiency observed is up to $35\%$ when multi-user sequencing is optimized, under given maximum transmit power and delay constraints.
In an alternate scenario, when the length of the transmission blocks is fixed and equal for each device, multi-user sequence optimization still provides a performance gain of up to $45\%$. However, the overall system performance degrades quite significantly.
The energy efficiency gain of multi-user sequence optimization is paramount under a stringent delay bound. Thus, multi-user sequence optimization makes the TDMA-based multi-user transmissions  more likely to be feasible in the lower latency regime subject to the given power constraints.

\appendices

\vspace{-0.25cm}
\section{Proof of Lemma 1}\label{A}
\vspace{-0.2cm}
It can be shown that the energy cost of the $i$th device, $E_i$, defined in \eqref{En}, is non-convex in $P_i$. By substitution of variable $Z_i = \ln \big(  1 {+} \kappa \frac {P_i |h_i|^2}{\Gamma \sigma^2 d^\alpha_i} \big)$ in \eqref{En}, $E_i$ can equivalently be expressed as
\vspace{-0.60cm}
\begin{equation}\label{En-Z}
  E_i \big(Z_i, D_{\textup{cp},i} \big) =  \tau D_i P_{\textup{cp}} \big( \big( D_i D^{-1}_{\textup{cp},i} \big)^\beta {-} 1 \big) +  D_{\textup{cp},i} b_i Z^{-1}_i  \big( \exp  ( Z_i ) {+} c_i  \big),
\end{equation}
\vspace{-0.2cm}
\noindent where $b_i{=}\frac{\Gamma \sigma^2 d^{\alpha}_i \ln(2)}{\mu B \kappa |h_i|^2}$, $c_i{=} \frac{\mu \kappa |h_i|^2 P_\textup{o}} {\Gamma \sigma^2 d^{\alpha}_i} {-} 1$.
Substituting $Z_i$ in constraints \eqref{opt-prob-T-1}, \eqref{opt-prob-d} and \eqref{opt-prob-g}  yields
\vspace{-0.60cm}
\begin{subequations}
\begin{alignat}{3}
& x_{1,i} \tau D_i \big( \big( D_i D^{-1}_{\textup{cp},i} \big)^\beta {-} 1 \big)    +   x_{1,i}  D_{\textup{cp},i} (B Z_i)^{-1} \leqslant T_1, \quad \forall \, i,      \label{opt-prob-T-1-2}\\
& \qquad \sum\nolimits_{n=2}^{N} x_{n,i}  D_{\textup{cp},i} (B Z_i)^{-1} \leqslant \sum\nolimits_{n=2}^{N} x_{n,i} T_n,  \quad \forall \, i,       \label{opt-prob-d-2}\\
& \qquad\qquad\qquad 0 \leqslant Z_i \leqslant Z_{\textup{max}}, \quad  \forall \, i,      \label{opt-prob-g-2}
\end{alignat}
\end{subequations}
\vspace{-0.15cm} \noindent where $Z_{\textup{max}} = \ln \big(  1 {+} \kappa \frac {P_{\textup{max}} |h_i|^2}{\Gamma \sigma^2 d^\alpha_i} \big)$.
Accordingly, problems \eqref{opt-prob-sum} and \eqref{opt-prob-equal} respectively become as
\begin{equation}\label{opt-prob-sum-2}
\begin{aligned}
\hspace{-1.5cm}\mathbf{\tilde{P}_{SUM}:} \quad &  \underset{Z_i, \, D_{\textup{cp},i}, \, T_n, \, \forall \, n,i }    {\textup{minimize}}
& & \quad  \sum\nolimits_{i=1}^{N} E_i (Z_i, D_{\textup{cp},i} ) \\
& \quad \textup{subject to}
& &   \quad \eqref{opt-prob-b}, \, \eqref{opt-prob-c}, \, \eqref{opt-prob-T-1-2}, \, \eqref{opt-prob-d-2}, \, \eqref{opt-prob-g-2}, \, \eqref{opt-prob-h}, \, \eqref{opt-prob-i},
\end{aligned}
\end{equation}
\vspace{-0.1cm}
\begin{equation}\label{opt-prob-equal-2}
\begin{aligned}
\hspace{-1.5cm}\mathbf{\tilde{P}_{MM}:} \quad &  \underset{Z_i, \, D_{\textup{cp},i}, \, T_n, \, \forall \, n, i }    {\textup{minimize}}
& & \quad  \underset{1\leqslant i \leqslant N}{\max} \big\{ E_i (Z_i, D_{\textup{cp},i} )  \big\} \\
&  \quad \textup{subject to}
& &   \quad \eqref{opt-prob-b}, \, \eqref{opt-prob-c}, \, \eqref{opt-prob-T-1-2}, \, \eqref{opt-prob-d-2}, \, \eqref{opt-prob-g-2}, \, \eqref{opt-prob-h}, \, \eqref{opt-prob-i}.
\end{aligned}
\end{equation}

Using basic calculus and with some algebraic manipulation, it can be shown that $E_i \big(Z_i, D_{\textup{cp},i} \big)$ in \eqref{En-Z} and constraint functions in \eqref{opt-prob-c}, \eqref{opt-prob-T-1-2} and \eqref{opt-prob-d-2}, respectively, are jointly convex in $Z_i$ and $D_{\textup{cp},i}, \, \forall \, i$. For brevity, we omit the detailed proof of this result here.
Also, because the sum of convex functions is convex and the maximum of the convex functions is also convex \cite{boyd2004}. $\sum_{i=1}^{N} E_i (Z_i, D_{\textup{cp},i} )$ and $\underset{1\leqslant i \leqslant N}{\max} \big\{ E_i (Z_i, D_{\textup{cp},i} )  \big\}$ both are jointly convex in $Z_i$ and $D_{\textup{cp},i}, \, \forall \, i$. Hence, for a given sequence, $\mathbf{\tilde{P}_{SUM}}$ and $\mathbf{\tilde{P}_{MM}}$ both are convex optimization problems.

Now consider problem  $\mathbf{\hat{P}_{PF}}$ in \eqref{opt-prob-propo-3}. It can be shown that the objective function in \eqref{opt-prob-propo-3} is jointly non-convex in $P_i$ and $D_{\textup{cp},i}, \, \forall \, i$. We propose substituting $ Z_i = \ln \big(  1 + \kappa \frac {P_i |h_i|^2}{\Gamma \sigma^2 d^\alpha_i} \big) $ and
$V_i = \ln \big( D_{\textup{cp},i} \big)$ in \eqref{En} to arrive at
\begin{equation}\label{En-ZV}
\log\Big(E_i \big(Z_i, V_i \big) \Big)= \log \Big( \tau D_i P_{\textup{cp}}  \Big( \frac {D^\beta_i} {\exp (\beta V_i)} {-} 1 \Big) + \frac{ b_i \exp ( V_i )} { Z_i } \Big( \exp\big( Z_i \big) {+} c_i  \Big) \Big).
\end{equation}

Substituting $Z_i$ and $V_i$ in the constraint functions \eqref{opt-prob-T-1}, \eqref{opt-prob-c}, \eqref{opt-prob-d}, \eqref{opt-prob-g} and \eqref{opt-prob-h} yields
\begin{subequations}
\begin{alignat}{3}
& x_{1,i} \tau D_i  \big( D^\beta_i \exp \big({-}\beta V_i\big) {-} 1 \big) + x_{1,i}  \exp \big(V_i\big) \ln(2) (B Z_i)^{-1} \leqslant T_1,  \quad \forall \, i,     \label{opt-prob-pf-T-1-2}\\
&\quad\sum\nolimits_{n=2}^{N} x_{n,i}\tau D_i\big( D^\beta_i \exp \big({-}\beta V_i\big){-}1 \big) \leqslant \sum\nolimits_{n=2}^{N} \sum\nolimits_{k=1}^{n-1} x_{n,i} T_k, \quad \forall \, i, \label{opt-prob-c-pf-2}\\
& \qquad \sum\nolimits_{n=2}^{N} x_{n,i} \exp \big(V_i\big) \ln(2)  (B Z_i)^{-1} \leqslant \sum\nolimits_{n=2}^{N} x_{n,i}T_n,   \quad \forall \, i,      \label{opt-prob-d-pf-2}\\
&  \qquad\qquad\qquad\qquad\quad 0 \leqslant Z_i \leqslant Z_{\textup{max}}, \quad  \forall \, i,           \label{opt-prob-g-pf-2}\\
& \qquad\qquad\qquad\qquad \ln(D_{\textup{min},i}) \leqslant V_i \leqslant \ln(D_i), \quad \forall \, i,          \label{opt-prob-h-pf-2}
\end{alignat}
\end{subequations}

For a known sequence, problem $\mathbf{\hat{P}_{PF}}$ in \eqref{opt-prob-propo-3} can equivalently be written as
\begin{equation}\label{opt-prob-propo-2}
\begin{aligned}
\hspace{-1.5cm}\mathbf{\tilde{P}_{PF}:} \quad &  \underset{Z_i, \, V_i, \, T_n, \, \forall \, n, i }    {\textup{minimize}}
& & \quad  \sum\nolimits_{i=1}^{N} \log \Big( E_i (Z_i, V_i ) \Big) \\
& \textup{subject to}
& & \quad    \eqref{opt-prob-b}, \, \eqref{opt-prob-pf-T-1-2}, \, \eqref{opt-prob-c-pf-2}, \, \eqref{opt-prob-d-pf-2}, \, \eqref{opt-prob-g-pf-2}, \, \eqref{opt-prob-h-pf-2}, \, \eqref{opt-prob-i}.
\end{aligned}
\end{equation}

It can also be shown that \eqref{En-ZV} is jointly convex in $Z_i$ and $V_i$. Since the sum of convex functions is convex \cite{boyd2004}, $\sum_{i=1}^{N} \log \big( E_i (Z_i, V_i) \big)$ is jointly convex in both $Z_i$ and $V_i, \, \forall \, i$.
Similarly, it can be shown that \eqref{opt-prob-c-pf-2} is convex in $V_i, \, \forall \, i$, \eqref{opt-prob-d-pf-2} is jointly convex in $Z_i$ and $V_i, \, \forall \, i$, and  \eqref{opt-prob-pf-T-1-2} is jointly convex in $Z_1$ and $V_1$. Hence, for a given sequence, $\mathbf{\tilde{P}_{PF}}$ is convex.

\vspace{-0.15cm}
\section{Proof of Lemma 3}\label{B}
\vspace{-0.05cm}
This proof is based on \cite{duy2018}, \cite{duy-36}, \cite{duy-25}. Let $\mathcal{S}^{(j{+}1)}=(Z^{(j{+}1)}_i, D^{(j{+}1)}_{\textup{cp},i}, T^{(j{+}1)}_n, x^{(j{+}1)}_{n,i}, \, \forall \, n,i)$ be the solution to problem \eqref{opt-prob-real-convex} at the $(j{+}1)$th iteration at a given point $x^{(j)}_{n,i}, \, \forall \, n,i,$, which yields
\vspace{-0.1cm}
\begin{equation*}
  \text{OF}^{(j{+}1)}\big( \mathcal{S}^{(j{+}1)} \big) = \sum\nolimits_{i=1}^{N} E^{(j{+}1)}_i (Z^{(j{+}1)}_i, D^{(j{+}1)}_{\textup{cp},i} )  {+} \Lambda  \sum\nolimits_{n=1}^{N} \sum\nolimits_{i=1}^{N} \Big( x^{(j{+}1)}_{n,i} \big( 1 {-} 2 x^{(j)}_{n,i} \big) {+} \big( x^{(j)}_{n,i} \big)^2  \Big) \\
\end{equation*}
\begin{equation}\label{lem-3-1}
  \leqslant
   \sum\nolimits_{i=1}^{N} E^{(j)}_i (Z^{(j)}_i, D^{(j)}_{\textup{cp},i} ) {+} \Lambda  \sum\nolimits_{n=1}^{N} \sum\nolimits_{i=1}^{N} \Big( x^{(j)}_{n,i} \big( 1 {-} 2 x^{(j{-}1)}_{n,i} \big) {+} \big( x^{(j{-}1)}_{n,i} \big)^2  \Big) = \text{OF}^{(j)} \big( \mathcal{S}^{(j)} \big)
\end{equation}


Thus, Algorithm 1 produces a monotone sequence of improved solutions, $\big\{ \text{OF}^{(j)} \big( \mathcal{S}^{(j)} \big) \big\}$, for problem \eqref{opt-prob-sum}. Moreover, $\big\{ \text{OF}^{(j)} \big( \mathcal{S}^{(j)} \big) \big\}$ is bounded by constraint functions \eqref{opt-prob-g}, \eqref{opt-prob-h} and \eqref{opt-prob-i}, and therefore convergence is guaranteed, i.e.,
$\Big( \text{OF}^{(j{+}1)}\big( \mathcal{S}^{(j{+}1)} \big) {-} \text{OF}^{(j)} \big( \mathcal{S}^{(j)} \big) \Big) \rightarrow 0$ as $j \rightarrow \infty$.

Following \cite{duy-36}, problem \eqref{opt-prob-sum}, which is solved by Algorithm~1, can be represented as
\begin{subequations}\label{lem-3-2}
\begin{alignat}{3}
  &\underset{\mathbf{s} }    {\textup{minimize}}
& & \quad  \varphi_\textup{o} ( \mathbf{s} ) \label{lem-3-2a} \\
&   \textup{subject to}
& &     \quad \psi_u        (\mathbf{s} ) \leqslant 0, \quad \forall \, u \, \in \, \{1,2,\cdot\cdot\cdot,U\}, \label{lem-3-2b}\\
& & & \quad  \varphi_v (\mathbf{s} ) \leqslant 0, \quad \forall \, v, \, \in \, \{1,2,\cdot\cdot\cdot,V\}, \label{lem-3-2c}
\end{alignat}
\end{subequations}
\noindent where $\mathbf{s}$ are design parameters, $\varphi_\textup{o}$ is the objective function, $\psi_u$ are convex constraints and $\varphi_v$ are non-convex constraints. The convex approximation of \eqref{opt-prob-sum} in \eqref{opt-prob-real-convex} can be rewritten as
\begin{subequations}\label{lem-3-3}
\begin{alignat}{3}
  &\underset{\mathbf{s} }    {\textup{minimize}}
& & \quad  \tilde{\varphi}_\textup{o} ( \mathbf{s}, \mathbf{s}^{(j)} ) \label{lem-3-3a} \\
&   \textup{subject to}
& &     \quad  \psi_u        (\mathbf{s}, \mathbf{s}^{(j)} ) \leqslant 0, \quad \forall \, u \, \in \, \{1,2,\cdot\cdot\cdot,U\}, \label{lem-3-3b}\\
& & & \quad  \tilde{\varphi}_v (\mathbf{s}, \mathbf{s}^{(j)} ) \leqslant 0, \quad \forall \, v, \, \in \, \{1,2,\cdot\cdot\cdot,V\}, \label{lem-3-3c}
\end{alignat}
\end{subequations}
\noindent where $\tilde{\varphi}_\textup{o}$ and $\tilde{\varphi}_v$ are the convex approximations of objective function, $\varphi_\textup{o}$, and the non-convex constraints, $\tilde{\varphi}_v$, respectively, at a given point $\mathbf{s}^{(j)}$. From \eqref{bi-convex-2}, \eqref{opt-prob-real-convex} and \eqref{lem-3-1}, we have
\begin{subequations}\label{lem-3-4}
\begin{alignat}{3}
    \varphi_v (\mathbf{s} )    ~\leqslant~&    \tilde{\varphi}_v (\mathbf{s}, \mathbf{s}^{(j)} ),  \quad \forall \, v,                \label{lem-3-4a} \\
     \varphi_v (\mathbf{s}^{(j)} )    ~=~&   \tilde{\varphi}_v (\mathbf{s}^{(j)}, \mathbf{s}^{(j)} ),   \quad \forall \, v,           \label{lem-3-4b}\\
    \nabla \varphi_v (\mathbf{s}^{(j)} )   ~=~&   \tilde{\varphi}_v (\mathbf{s}^{(j)}, \mathbf{s}^{(j)} ), \quad \forall \, v.   \label{lem-3-4c}
\end{alignat}
\end{subequations}

Let $\mathbf{s}^{(j)}$ be the solution for Algorithm~1 at convergence. It is the optimal and the Fritz John point for problem \eqref{lem-3-3} satisfying the following conditions \cite{duy-25}
\begin{subequations}\label{lem-3-5}
\begin{alignat}{3}
   &   \lambda_\textup{o} \nabla \tilde{\varphi}_\textup{o} (\mathbf{s}^{(j)}, \mathbf{s}^{(j)} )
   +    \sum\nolimits_{u=1}^{U}   \lambda_u \nabla \psi_u  ( \mathbf{s}^{(j)} )
   +    \sum\nolimits_{v=1}^{V}   \lambda_v \nabla \tilde{\varphi}_v (\mathbf{s}^{(j)}, \mathbf{s}^{(j)} ) = 0,                 \label{lem-3-5a} \\
     &  \qquad\qquad\qquad\qquad\qquad\lambda_u  \psi_u  ( \mathbf{s}^{(j)} )  = 0, \quad \forall \, u,             \label{lem-3-5b}\\
    &   \qquad\qquad\qquad\qquad\quad\, \lambda_v \tilde{\varphi}_v (\mathbf{s}^{(j)}, \mathbf{s}^{(j)} ) = 0, \quad \forall \, v,   \label{lem-3-5c}
\end{alignat}
\end{subequations}
\vspace{-0.5cm}
\noindent where $ \lambda_u$ and $ \lambda_v$ are Lagrange variables for the $u$th and the $v$th convex and non-convex constraints, respectively.
Substituting  \eqref{lem-3-4b} and \eqref{lem-3-4c} in \eqref{lem-3-5} yields
\begin{subequations}\label{lem-3-6}
\begin{alignat}{3}
   &   \lambda_\textup{o} \nabla \varphi_\textup{o} (\mathbf{s}^{(j)} )
   +    \sum\nolimits_{u=1}^{U}   \lambda_u \nabla \psi_u  ( \mathbf{s}^{(j)} )
   +    \sum\nolimits_{v=1}^{V}   \lambda_v \nabla \varphi_v ( \mathbf{s}^{(j)} ) = 0,                 \label{lem-3-6a} \\
     &  \qquad\qquad\qquad\qquad\quad\,  \lambda_u  \psi_u  ( \mathbf{s}^{(j)} )  = 0, \quad \forall \, u,             \label{lem-3-6b}\\
    &   \qquad\qquad\qquad\qquad\quad\, \lambda_v \varphi_v (\mathbf{s}^{(j)} ) = 0, \quad \forall \, v.   \label{lem-3-6c}
\end{alignat}
\end{subequations}
\noindent The above results hold  and thus satisfy the conditions \eqref{lem-3-5}, implying that $\mathbf{s}^{(j)}$ is a Fritz John solution of \eqref{lem-3-2}. Since \eqref{lem-3-2} represents \eqref{opt-prob-sum} the same conclusion can be drawn for \eqref{opt-prob-sum}. \qed 

\section{Optimization Problems for the Benchmark Scheme}\label{C}

For the benchmark scheme, the problems for the considered objectives can be expressed as
\begin{subequations}\label{opt-prob-b11-sum}
\begin{alignat}{2}
\hspace{1.7cm}\mathbf{P^{b}_{SUM}:} \quad &  \underset{  Z_i, \, T_n, \, \forall \, n, i } {\textup{minimize}}
& & \quad  \sum\nolimits_{i=1}^{N} \Big( D_{i} b_i Z^{-1}_i  \big( \exp ( Z_i ) + c_i  \big)  \Big)    \label{b1`-sum-a} \\
& \textup{subject to}
& &    \quad \eqref{opt-prob-b},  \eqref{opt-prob-i},  \eqref{b1-sum-d} ,                         \nonumber\\
& & &  \quad\sum\nolimits_{n=1}^{N} x_{n,i}  D_i (B Z_i)^{-1} \leqslant \sum\nolimits_{n=1}^{N} x_{n,i} T_n,  \quad \forall \, i,      \label{b11-sum-c}
\end{alignat}
\end{subequations}
\begin{equation}\label{opt-prob-b11-equal}
\begin{aligned}
\hspace{-1.52cm}\mathbf{P^{b}_{MM}:} \quad &  \underset{  Z_i, \, T_n, \, \forall \, n, i } {\textup{minimize}}
& &   \underset{1\leqslant i \leqslant N}{\max} \Big\{ D_{i} b_i Z^{-1}_i  \big( \exp ( Z_i ) + c_i   \big)  \Big\}  \\
& \textup{subject to}
& &    \eqref{opt-prob-b},  \eqref{opt-prob-i},  \eqref{b1-sum-d} ,  \eqref{b11-sum-c},
\end{aligned}
\end{equation}
\begin{equation}\label{opt-prob-b11-log}
\begin{aligned}
\hspace{-1.0cm}\mathbf{P^{b}_{PF}:} \quad &  \underset{  Z_i, \, T_n, \, \forall \, n, i } {\textup{minimize}}
& &   \sum_{i=1}^{N} \log \Big( D_{i} b_i Z^{-1}_i  \big( \exp ( Z_i ) + c_i   \big)  \Big)     \\
& \textup{subject to}
& &     \eqref{opt-prob-b},   \eqref{opt-prob-i}, \eqref{b1-sum-d} ,  \eqref{b11-sum-c}, .
\end{aligned}
\end{equation}

\section{Optimization Problems for Case 1}\label{D}

For Case 1, the problems for the considered objectives can be expressed as
\begin{subequations}\label{opt-prob-b3-sum}
\begin{alignat}{2}
\mathbf{P^{c1}_{SUM}:} \quad &  \underset{  Z_i, \, D_{\textup{cp},i}, \,  \forall \, i } {\textup{minimize}}
& & \quad  \sum\nolimits_{i=1}^{N} \Big(  \tau D_n P_{\textup{cp}} \big( (D_n D^{-1}_{\textup{cp},n} )^\beta {-} 1 \big)
{+} D_{\textup{cp},n} b_n Z^{-1}_n  \big( \exp ( Z_n ) {+} c_n  \big)  \Big)    \label{b3-sum-a} \\
& \textup{subject to}
& & \quad\eqref{opt-prob-h}, \eqref{b1-sum-d},  \nonumber \\
& & & \quad x_{1,i} \tau D_i \Big( \big(D_i D^{-1}_{\textup{cp},i} \big)^\beta {-} 1 \Big)    +   x_{1,i}  D_{\textup{cp},i}  (B Z_i)^{-1} \leqslant T_\textup{frame} N^{-1}, \quad \forall \, i,   \label{b3-T-1} \\
& &  & \quad \sum\nolimits_{n=2}^{N} x_{n,i} \tau D_i \Big( \big( \frac{D_i} {D_{\textup{cp},i}} \big)^\beta {-} 1 \Big) \leqslant \sum\nolimits_{n=2}^{N} \sum\nolimits_{k=1}^{n-1} x_{n,i} \frac{T_\textup{frame}}{ N},  \quad \forall \, i, \label{b3-sum-b}\\
& & &  \quad \sum\nolimits_{n=2}^{N} x_{n,i} D_{\textup{cp},i} (B Z_i)^{-1} \leqslant \sum\nolimits_{n=2}^{N} x_{n,i} T_\textup{frame} N^{-1},  \quad \forall \, i,      \label{b3-sum-c}
\end{alignat}
\end{subequations}
\begin{equation}\label{opt-prob-b3-equal}
\begin{aligned}
\hspace{-3.5cm}\mathbf{P^{c1}_{MM}:} \quad &  \underset{  Z_i, \, D_{\textup{cp},i}, \, \forall \, i } {\textup{minimize}}
& &   \underset{1\leqslant i \leqslant N}{\max} \Big\{ \tau D_i P_{\textup{cp}} \Big( \big(D_i D^{-1}_{\textup{cp},i} \big)^\beta {-} 1 \Big)
{+} D_{\textup{cp},i} b_i Z^{-1}_i  \big( \exp ( Z_i ) {+} c_i  \big) \Big\}  \\
& \textup{subject to}
& &    \eqref{opt-prob-h}, \eqref{b1-sum-d},  \textup{\cref{b3-T-1,b3-sum-b,b3-sum-c}},
\end{aligned}
\end{equation}
\begin{subequations}\label{opt-prob-b3-log}
\begin{alignat}{2}
\hspace{-0.01cm}\mathbf{P^{c1}_{PF}:} \quad &  \underset{  Z_i, \, V_i,  \, \forall \, i } {\textup{minimize}}
& &   \quad \sum\nolimits_{i=1}^{N} \log \Big( \tau D_i P_{\textup{cp}}  \Big( \frac {D^\beta_i} {\exp (\beta V_i)} {-} 1 \Big) {+} \frac{ b_i \exp \big( V_i \big)} { Z_i } \Big( \exp\big( Z_i \big) {+} c_i  \Big) \Big)      \\
& \textup{subject to}
& &  \quad  \eqref{b1-sum-d},  \eqref{b2-propo-d}, \nonumber  \\
& & & \quad x_{1,i} \tau D_i \Big( \frac {D^\beta_1} {\exp (\beta V_1)} {-} 1 \Big)    +   x_{1,i}  \frac{\exp (V_1) \ln(2)}  {B Z_i} \leqslant \frac{T_\textup{frame}}{N}, \quad \forall \, i,  \label{b3-pf-T-1} \\
%
& & & \quad \sum\nolimits_{n=2}^{N} x_{n,i} \tau D_i \Big( \big( \frac{D^{\beta}_i} {\exp (\beta V_1)} \big) {-} 1 \Big) \leqslant \sum\nolimits_{n=2}^{N} \sum\nolimits_{k=1}^{n-1} x_{n,i} \frac{T_\textup{frame}}{ N},  \quad \forall \, i,  \label{b3-pf-T-2} \\
& & & \quad  \sum\nolimits_{n=2}^{N} x_{n,i} \exp (V_n) \ln(2) (B Z_i)^{-1} \leqslant \sum\nolimits_{n=2}^{N} x_{n,i} T_\textup{frame} N^{-1},  \quad \forall \, i,  \label{b3-pf-T-3}
\end{alignat}
\end{subequations}

\section{Optimization Problems for Case 2}\label{E}

For Case 2, the problems for the considered objectives can be expressed as
\begin{equation}\label{opt-prob-b4-sum}
\begin{aligned}
\hspace{-1.10cm}\mathbf{P^{c2}_{SUM}:} \quad &  \underset{\substack{  Z_i, \, D_{\textup{cp},i}, \\ x_{n,i},  \, \forall \, n, i }  }    {\textup{minimize}}
& & \quad  \sum\nolimits_{i=1}^{N} \Big(  \tau D_i P_{\textup{cp}} \Big( \big(D_i D^{-1}_{\textup{cp},i} \big)^\beta {-} 1 \Big)
{+} D_{\textup{cp},i} b_i Z^{-1}_i  ( \exp ( Z_i )  {+} c_i  )  \Big)    \\
& \textup{subject to}
& &  \quad  \eqref{opt-prob-e}, \eqref{opt-prob-f},  \eqref{opt-prob-h}, \eqref{opt-prob-j},  \eqref{b1-sum-d}, \textup{\cref{b3-T-1,b3-sum-b,b3-sum-c}},
\end{aligned}
\end{equation}
\begin{equation}\label{opt-prob-b4-equal}
\begin{aligned}
\hspace{-1.80cm}\mathbf{P^{c2}_{MM}:} \quad &  \underset{\substack{ Z_i, \, D_{\textup{cp},i}, \\  x_{n,i}, \, \forall \, n, i }  }    {\textup{minimize}}
& &   \underset{1\leqslant i \leqslant N}{\max} \Big\{  \tau D_i P_{\textup{cp}} \Big( \big(D_i D^{-1}_{\textup{cp},i} \big)^\beta {-} 1 \Big)
{+} D_{\textup{cp},i} b_i Z^{-1}_i  ( \exp ( Z_i )  {+} c_i  ) \Big\}  \\
& \textup{subject to}
& &   \eqref{opt-prob-e}, \eqref{opt-prob-f},  \eqref{opt-prob-h}, \eqref{opt-prob-j},  \eqref{b1-sum-d}, \textup{\cref{b3-T-1,b3-sum-b,b3-sum-c}},
\end{aligned}
\end{equation}
\begin{equation}\label{opt-prob-b4-log}
\begin{aligned}
\hspace{0.37cm}\mathbf{P^{c2}_{PF}:} \quad &  \underset{\substack{ Z_i, \, V_i, \\ x_{n,i}, \, \forall \, n, i }  }    {\textup{minimize}}
& &   \sum\nolimits_{i=1}^{N} \log \Big(  \tau D_i P_{\textup{cp}}  \Big( \frac {D^\beta_i} {\exp (\beta V_i)} {-} 1 \Big) {+} \frac{ b_i \exp \big( V_i \big)} { Z_i } \Big( \exp\big( Z_i \big) {+} c_i  \Big) \Big)     \\
& \textup{subject to}
& &  \eqref{opt-prob-e}, \eqref{opt-prob-f}, \eqref{opt-prob-j},  \eqref{b1-sum-d}, \, \eqref{b2-propo-d}, \, \textup{\cref{b3-pf-T-1,b3-pf-T-2,b3-pf-T-3}}.
\end{aligned}
\end{equation}

\end{document}